\documentclass[twocolumn]{aastex631}
\usepackage{graphicx}

\newcommand{\tnm}[1]{\tablenotemark{#1}}

\newcommand{\perpixel}{\ensuremath{\mathrm{pixel}^{-1}}}
\newcommand{\peryr}{\ensuremath{\mathrm{yr}^{-1}}}
\newcommand{\kms}{\mbox{km\ s$^{-1}$}}
\newcommand{\kmsMpc}{\kms~\mbox{Mpc}$^{-1}$}
\newcommand{\MUV}{\ensuremath{M_{\rm UV}}}
\newcommand{\Msun}{\ensuremath{M_{\odot}}}

\newcommand{\Zsun}{\ensuremath{Z_{\odot}}}

\newcommand{\Mstar}{\ensuremath{M_{*}}}

\newcommand{\logmstar}{\ensuremath{\log M_{\star}/\Msun}}

\newcommand{\Ho}{\ensuremath{H_{0}}}
\newcommand{\Om}{\Omega_{\rm m}}
\newcommand{\OL}{\Omega_{\Lambda}}

\newcommand{\mAB}{\ensuremath{m_{\rm AB}}}
\newcommand{\MAB}{\ensuremath{M_{\rm AB}}}
\newcommand{\zphot}{\ensuremath{z_{\rm phot}}}
\newcommand{\zspec}{\ensuremath{z_{\rm spec}}}

\newcommand{\tform}{\ensuremath{t_{\rm form}}}

\newcommand{\fnu}{\ensuremath{f_{\nu}}}

\newcommand{\wl}{\mbox{$\lambda$}}
\newcommand{\ionrm}[1]{\mbox{\small\sc{\romannumeral #1}}}
\newcommand{\forb}[2]{\mbox{[#1~\ionrm{#2}]}}

\newcommand{\forbww}[4]{\mbox{[#1~\ionrm{#2}]~\wl\wl#3,~#4}}

\newcommand{\OIIww}{\forbww{O}{2}{3726}{3729}}

\newcommand{\OIIIfir}{\forb{O}{3} 88~$\mu$m}
\newcommand{\Ha}{\mbox{H\ensuremath{\alpha}}}

\newcommand{\HII}{H\,\textsc{ii}}
\newcommand{\HeIIw}{He~\ionrm{2}~\wl1640~\AA}
\newcommand{\Lya}{Ly\ensuremath{\alpha}}

\newcommand{\grizli}{\textsc{grizli}}
\newcommand{\eazypy}{\textsc{eazypy}}
\newcommand{\bagpipes}{\textsc{bagpipes}}

\newcommand{\photutils}{\textsc{photutils}}

\newcommand{\SEP}{\texttt{SEP}}
\newcommand{\astropy}{\textsc{astropy}}
\newcommand{\astrodrizzle}{\textsc{astrodrizzle}}
\newcommand{\multinest}{\textsc{multinest}}
\newcommand{\cloudy}{\textsc{Cloudy}}

\newcommand{\lenstool}{\texttt{LENSTOOL}}
\newcommand{\lta}{\textsc{LENSTOOL-A}}
\newcommand{\ltb}{\textsc{LENSTOOL-B}}
\newcommand{\glafic}{\textsc{Glafic}}
\newcommand{\wslap}{\textsc{WSLAP+}}

\newcommand{\SPT}{SPT-CL~J0615$-$5746}
\newcommand{\sptarc}{SPT0615-JD}
\newcommand{\arc}{Cosmic Gems Arc}

\accepted{2025 July 29, ApJ}

\begin{document}

\title{Unveiling the Cosmic Gems Arc at $z\sim10$ with JWST NIRCam}

\correspondingauthor{Larry D. Bradley}
\email{lbradley@stsci.edu}

\author[0000-0002-7908-9284]{Larry D. Bradley}
\affiliation{Space Telescope Science Institute (STScI), 3700 San Martin Drive, Baltimore, MD 21218, USA}

\author[0000-0002-8192-8091]{Angela Adamo}
\affiliation{The Oskar Klein Centre, Department of Astronomy, Stockholm University, AlbaNova, SE-106 91 Stockholm, Sweden}

\author[0000-0002-5057-135X]{Eros Vanzella}
\affiliation{INAF--OAS, Osservatorio di Astrofisica e Scienza dello Spazio di Bologna, via Gobetti 93/3, I-40129 Bologna, Italy}

\author[0000-0002-7559-0864]{Keren Sharon}
\affiliation{Department of Astronomy, University of Michigan, 1085 S. University Ave, Ann Arbor, MI 48109, USA}

\author[0000-0003-2680-005X]{Gabriel Brammer}
\affiliation{Cosmic Dawn Center (DAWN), Copenhagen, Denmark}
\affiliation{Niels Bohr Institute, University of Copenhagen, Jagtvej 128, DK-2200 N, Copenhagen, Denmark}

\author[0000-0001-7410-7669]{Dan Coe}
\affiliation{Space Telescope Science Institute (STScI), 3700 San Martin Drive, Baltimore, MD 21218, USA}
\affiliation{Association of Universities for Research in Astronomy (AURA) for the European Space Agency (ESA), STScI, Baltimore, MD, USA}
\affiliation{Center for Astrophysical Sciences, Department of Physics and Astronomy, The Johns Hopkins University, 3400 N Charles St. Baltimore, MD 21218, USA}

\author[0000-0001-9065-3926]{Jose M. Diego}
\affiliation{Instituto de F\'isica de Cantabria (CSIC-UC). Avda. Los Castros s/n. 39005 Santander, Spain}

\author[0000-0002-5588-9156]{Vasily Kokorev}
\affiliation{Department of Astronomy, The University of Texas at Austin, Austin, TX 78712, USA}

\author[0000-0003-3266-2001]{Guillaume Mahler}
\affiliation{Institute for Computational Cosmology, Durham University, South Road, Durham DH1 3LE, UK}
\affiliation{Centre for Extragalactic Astronomy, Durham University, South Road, Durham DH1 3LE, UK}
\affiliation{STAR Institute, Quartier Agora - All\'{e}e du six Ao\^{u}t, 19c B-4000 Li\`{e}ge, Belgium}

\author[0000-0003-3484-399X]{Masamune Oguri}
\affiliation{Center for Frontier Science, Chiba University, 1-33 Yayoi-cho, Inage-ku, Chiba 263-8522, Japan}
\affiliation{Department of Physics, Graduate School of Science, Chiba University, 1-33 Yayoi-Cho, Inage-Ku, Chiba 263-8522, Japan}

\author[0000-0002-5258-8761]{Abdurro'uf}
\affiliation{Center for Astrophysical Sciences, Department of Physics and Astronomy, The Johns Hopkins University, 3400 N Charles St. Baltimore, MD 21218, USA}
\affiliation{Space Telescope Science Institute (STScI), 3700 San Martin Drive, Baltimore, MD 21218, USA}

\author[0000-0003-0883-2226]{Rachana Bhatawdekar}
\affiliation{European Space Agency (ESA), European Space Astronomy Centre (ESAC), Camino Bajo del Castillo s/n, 28692 Villanueva de la Cañada, Madrid, Spain}

\author[0000-0001-8415-7547]{Lise Christensen}
\affiliation{Cosmic Dawn Center (DAWN), Copenhagen, Denmark}
\affiliation{Niels Bohr Institute, University of Copenhagen, Jagtvej 128, DK-2200 N, Copenhagen, Denmark}

\author[0000-0001-7201-5066]{Seiji Fujimoto}\altaffiliation{Hubble Fellow}
\affiliation{David A. Dunlap Department of Astronomy and Astrophysics, University of Toronto, 50 St. George Street, Toronto, Ontario, M5S 3H4, Canada}
\affiliation{Dunlap Institute for Astronomy and Astrophysics, 50 St. George Street, Toronto, Ontario, M5S 3H4, Canada}

\author[0000-0002-0898-4038]{Takuya Hashimoto}
\affiliation{Graduate School of Pure and Applied Sciences, University of Tsukuba, 1-1-1 Tennodai, Tsukuba, Ibaraki 305-8571, Japan}
\affiliation{Tomonaga Center for the History of the Universe, University of Tsukuba, 1-1-1 Tennodai, Tsukuba, Ibaraki 305-8571, Japan}

\author[0000-0003-4512-8705]{Tiger Y.-Y Hsiao}
\affiliation{Center for Astrophysical Sciences, Department of Physics and Astronomy, The Johns Hopkins University, 3400 N Charles St. Baltimore, MD 21218, USA}

\author[0000-0002-7779-8677]{Akio K. Inoue}
\affiliation{Department of Physics, School of Advanced Science and Engineering, Faculty of Science and Engineering, Waseda University, 3-4-1 Okubo, Shinjuku, Tokyo 169-8555, Japan}
\affiliation{Waseda Research Institute for Science and Engineering, Faculty of Science and Engineering, Waseda University, 3-4-1 Okubo, Shinjuku, Tokyo 169-8555, Japan}

\author[0000-0002-6090-2853]{Yolanda Jim\'enez-Teja}
\affiliation{Instituto de Astrof\'isica de Andaluc\'ia--CSIC, Glorieta de la Astronom\'ia s/n, E--18008 Granada, Spain}
\affiliation{Observat\'orio Nacional, Rua General Jos\'e Cristino, 77 - Bairro Imperial de S\~ao Crist\'ov\~ao, Rio de Janeiro, 20921-400, Brazil}

\author[0000-0003-1427-2456]{Matteo Messa}
\affiliation{INAF--OAS, Osservatorio di Astrofisica e Scienza dello Spazio di Bologna, via Gobetti 93/3, I-40129 Bologna, Italy}

\author[0000-0002-5222-5717]{Colin Norman}
\affiliation{Department of Physics and Astronomy, The Johns Hopkins University, 3400 N Charles St. Baltimore, MD 21218, USA}
\affiliation{Space Telescope Science Institute (STScI), 3700 San Martin Drive, Baltimore, MD 21218, USA}

\author[0000-0003-4223-7324]{Massimo Ricotti}
\affiliation{Department of Astronomy, University of Maryland, College Park, 20742, USA}

\author[0000-0003-4807-8117]{Yoichi Tamura}
\affiliation{Department of Physics, Graduate School of Science, Nagoya University Furo, Chikusa, Nagoya, Aichi 464-8602, Japan}

\author[0000-0001-8156-6281]{Rogier A. Windhorst}
\affiliation{School of Earth and Space Exploration, Arizona State University, Tempe, AZ 85287-1404, USA}

\author[0000-0002-9217-7051]{Xinfeng Xu}
\affiliation{Department of Physics and Astronomy, Northwestern University, 2145 Sheridan Road, Evanston, IL, 60208, USA}
\affiliation{Center for Interdisciplinary Exploration and Research in Astrophysics (CIERA), Northwestern University, 1800 Sherman Avenue, Evanston, IL, 60201, USA}

\author[0000-0002-0350-4488]{Adi Zitrin}
\affiliation{Department of Physics, Ben-Gurion University of the Negev, P.O. Box 653, Be'er-Sheva 84105, Israel}



\begin{abstract}

We present recent JWST NIRCam imaging observations of \sptarc\ (also
known as the \arc), lensed by the galaxy cluster \SPT. The 5\arcsec long
arc is the most highly magnified $z > 10$ galaxy known. It straddles
the lensing critical curve and reveals five star clusters with radii
of $\sim 1$~pc or less. We measure the full arc to have F200W 24.5 AB
mag, consisting of two mirror images, each 25.3 AB mag with a median
magnification of $\mu \sim 60_{-8}^{+17}$ (delensed 29.7 AB mag, $\MUV
= -17.8$). The galaxy has an extremely strong Lyman break F115W$-$F200W
$>3.2$ mag ($2\sigma$ lower limit), is undetected in all bluer filters
($< 2\sigma$), and has a very blue continuum slope redward of the break
($\beta = -2.7 \pm 0.1$). This results in a photometric redshift $\zphot
= 10.2 \pm 0.2$ (95\% confidence) with no significant likelihood below
$z < 9.8$. Based on spectral energy distribution fitting to the total
photometry, we estimate an intrinsic stellar mass of $\Mstar \sim 2.4 -
5.6 \times 10^{7} \Msun$, young mass-weighted age of $\sim 21 - 79$~Myr,
low dust content ($A_V < 0.15$), and a low metallicity of $\lesssim
1\%~\Zsun$. We identify a fainter third counterimage candidate within
2\farcs2 of the predicted position, lensed to AB mag 28.4 and magnified
by $\mu \sim 2$, suggesting the fold arc may only show $\sim60$\% of the
galaxy. \sptarc\ is a unique laboratory to study star clusters observed
within a galaxy just 460~Myr after the Big Bang.

\end{abstract}



\section{Introduction} \label{sec:intro}

The James Webb Space Telescope (JWST) was designed to peer into the
distant Universe and study galaxies near the beginning of time.
Early in its mission, JWST observations have already prompted us to
reevaluate our understanding of the first phases of galaxy build-up,
eventually leading to the reionization of the Universe. At early
cosmic times and throughout the reionization era, galaxies appear to
experience rapid starburst phases \citep{Endsley2023, Boyett2024} and
metal enrichment \citep{Curti2024}, merger events \citep{Asada2023,
Hsiao2023}, and harbor massive stars producing extreme ionization
\citep[e.g.,][]{Atek2023, Matthee2023}. At redshift $z > 10$ ($t_{\rm
Universe} < 470$~Myr), early galaxies appear to be more luminous than
expected, suggesting conditions for which we have not accounted.
Several explanations have so far been explored, including significantly
higher star formation efficiency during their early assembly stages,
a top-heavy initial mass function (IMF), and stochastic burst events
\citep[e.g,][among many others]{Adams2023, Harikane2023, McLeod2023,
Finkelstein2024}.

While JWST has already discovered many galaxies at $z > 9$
\citep[e.g.,][]{Castellano2022, Finkelstein2022, Naidu2022a, Adams2023,
Atek2023, Bradley2023, Donnan2023, Harikane2023, Castellano2024,
Finkelstein2024, Hainline2024, Robertson2024, Zavala2025}, most are too
faint and small to be studied in detail. Most of these early galaxies
will remain unresolved, with their stellar populations only inferred
and never observed directly. Highly lensed and spatially resolved early
galaxies \citep[e.g.,][]{Bradley2023, Hsiao2023, Roberts-Borsani2023,
Stiavelli2023, Vanzella2023, Bradac2024} will provide the only chance to
directly study the engines that reionized the Universe.

The combined powers of JWST and gravitational lensing have revealed
small star clusters in a precious few highly magnified distant
galaxies at $4 < z < 8$ \citep[e.g.,][]{Mowla2022, Claeyssens2023,
Vanzella2022_glass, Vanzella2023, Mowla2024}. Star clusters with
radii as small as $\sim 1$~pc were recently discovered in a highly
magnified, distant galaxy at $\zspec = 5.93$, dubbed the ``Sunrise Arc''
\citep{Vanzella2023}. Some of these are observed to be young massive
star clusters (YSCs; a few Myr old) with intense ionizing emission,
while others are somewhat older (a few hundred Myr) and already
gravitationally bound \citep{Vanzella2023}.

\sptarc\ (also known as the \arc), discovered by \cite{Salmon2018}
in the Reionization Lensing Cluster Survey (RELICS) Hubble Treasury
program \citep{Coe2019}, holds the record for being the most highly
magnified and second-brightest galaxy known at $z\gtrsim10$. Magnified
to 24.5~AB mag, SPT0615-JD is several magnitudes brighter than most new
$z\gtrsim10$ candidates discovered by JWST. Hubble Space Telescope (HST)
imaging at 1.6~$\micron$ (rest-frame UV at $\sim$1500~\AA) identified
\sptarc\ as a 2\farcs5 long arc \citep{Salmon2018} and revealed small
structures with radii of $25 - 70$~pc \citep{Welch2023}. The superb
resolution offered by NIRCam has enabled us to resolve the \arc\ into
five YSCs, which dominate the light of the galaxy \citep{Adamo2024}. The
combination of magnified brightness and resolution makes the \arc\ a
unique laboratory to conduct spatially resolved studies with JWST, not
possible in any other galaxy at this distance.

In this paper, we present JWST NIRCam imaging observations of the \SPT\
galaxy cluster and the \arc\ (\sptarc), and the redshift estimate
and the derived physical properties of the latter. We describe the
observations in Section~\ref{sec:obs}. Section~\ref{sec:methods}
presents the data reduction, photometric catalogs, and photometric
redshifts. Section~\ref{sec:lens_models} presents the lens models of
the foreground galaxy cluster. Section~\ref{sec:results} presents the
results and discussion. Section~\ref{sec:conclusions} summarizes the
results and conclusions. We use the absolute bolometric (AB) magnitude
system, $\mAB = 31.4 - 2.5 \log(\fnu\ / {\rm nJy})$ \citep{Oke1974,
OkeGunn1983}. Where needed, we adopt a concordance cosmology with $\Ho
= 70$~\kmsMpc, $\Om = 0.3$, and $\OL = 0.7$, for which $1\arcsec \sim
4.1$~kpc at $z = 10.2$. All photometric redshift uncertainties are given
at the 95\% confidence level.

\begin{figure*}[t!]
\plotone{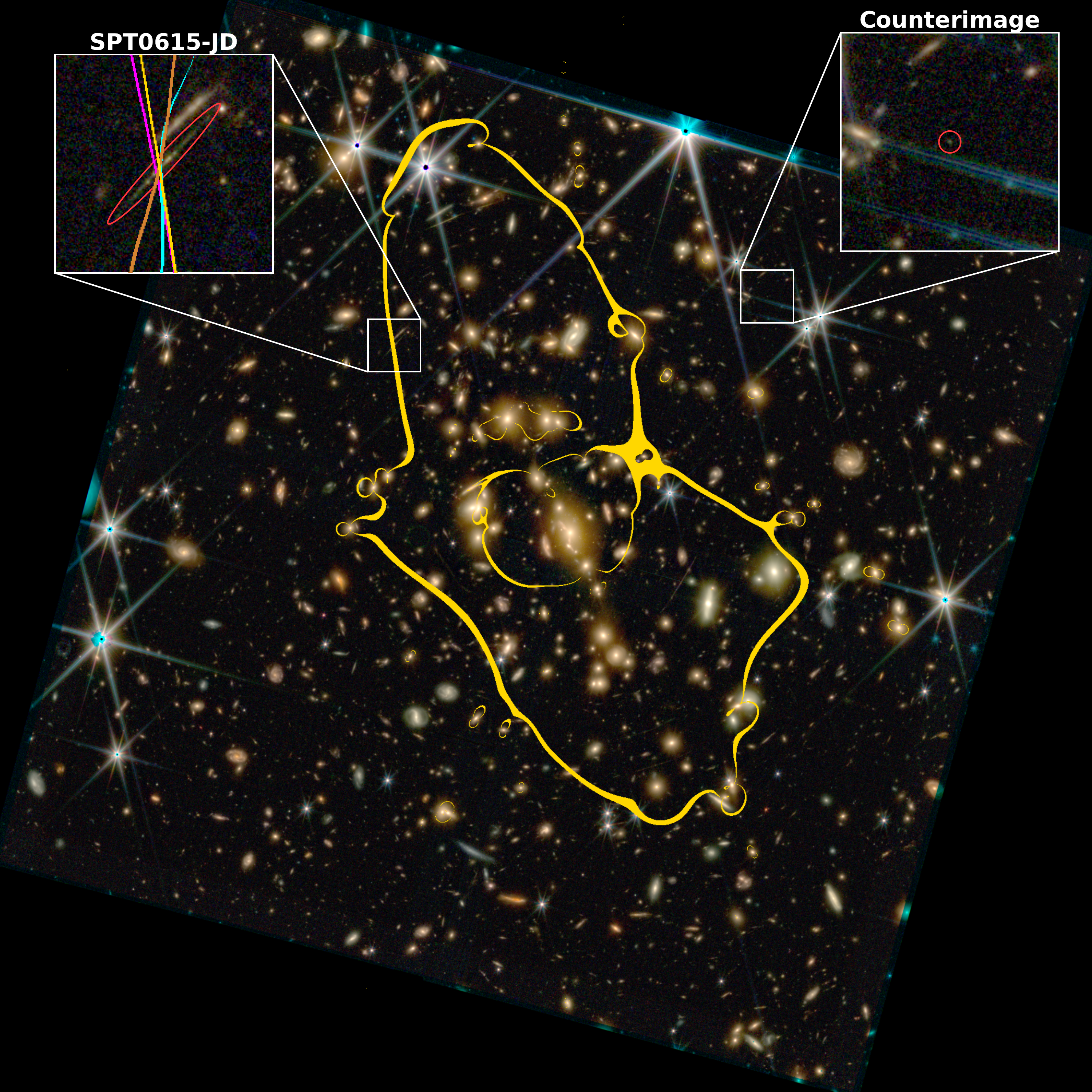}
\caption{
JWST NIRCam color image of the \SPT\ cluster field (red: F115W + F150W,
green: F200W + F277W, blue: F356W + F444W). The field of view is $\sim
2\farcm3 \times 2\farcm3$ and the image is shown with north up and east
left. The $z = 10.2$ critical curve of our fiducial \lta\ model (see
Section~\ref{sec:lens_models}) is shown in gold. The location of the
\arc\ is shown in the left-hand open white box, with a zoomed inset
figure ($8\arcsec \times 8\arcsec$) outlining the galaxy with a red
ellipse. The $z = 10.2$ critical curves of the \lta\ (gold), \ltb\ (dark
orange), \glafic\ (cyan), and \wslap\ (magenta) lens models (described
in Section~\ref{sec:lens_models}) bisect the \arc. The right-hand white
box and zoomed inset ($8\arcsec \times 8\arcsec$) shows the candidate
counterimage of the arc, which is located near (within 2\farcs2) the
position predicted by the lens models.
\label{fig:cluster}
}
\end{figure*}

\begin{figure*}[t!]
\plotone{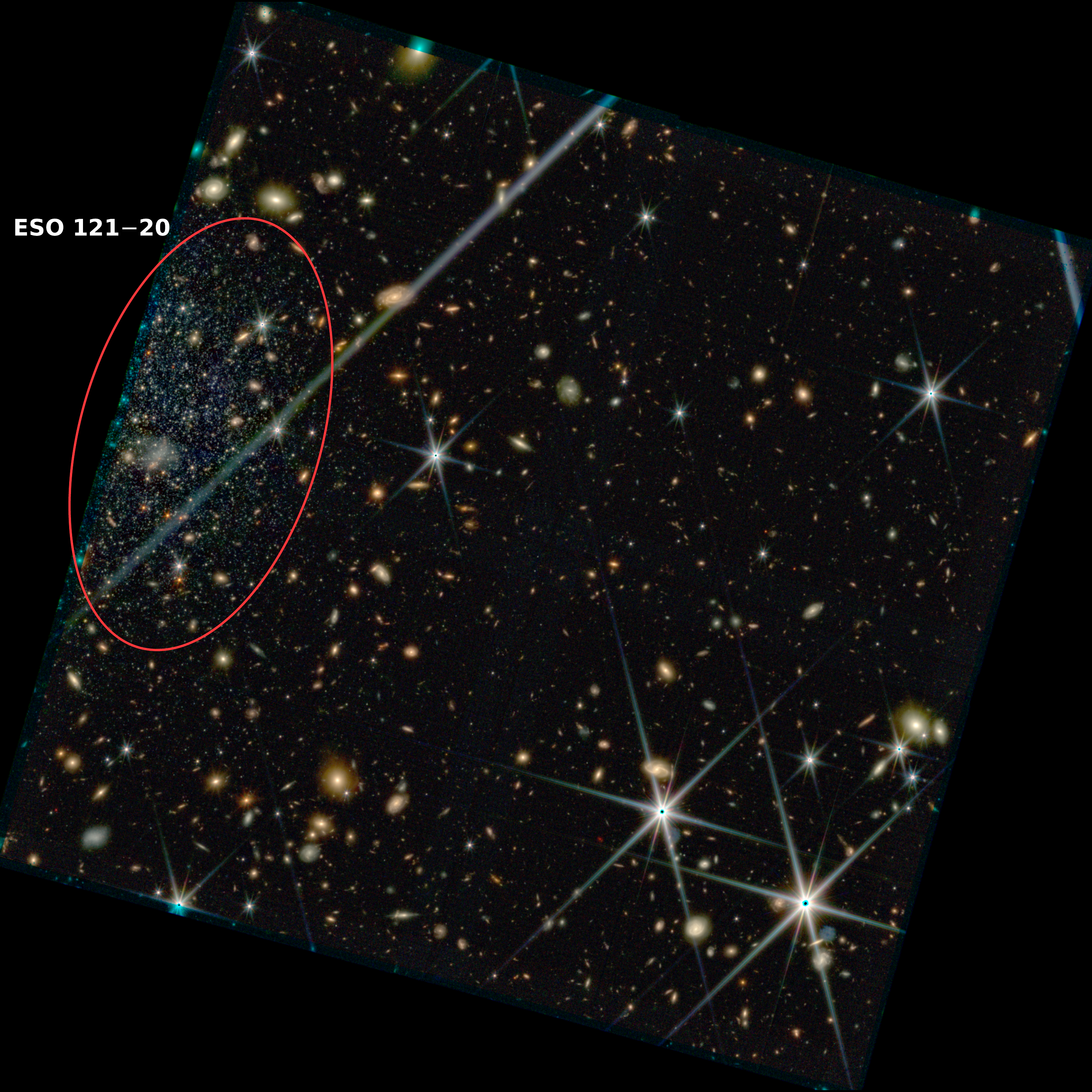}
\caption{
JWST NIRCam color image of the \SPT\ parallel field, located
north-northwest of the cluster field (red: F115W + F150W, green: F200W +
F277W, blue: F356W + F444W). The field of view is $\sim 2\farcm3 \times
2\farcm3$ and the image is shown with north up and east left. The field
partially includes ESO~121$-$20 (outlined in red), an isolated dwarf
irregular galaxy at a distance of 6.05~Mpc \citep{Karachentsev2006}.
\label{fig:parallel}
}
\end{figure*}

\section{Observations} \label{sec:obs}

\subsection{JWST Data}

We obtained JWST NIRCam imaging of the galaxy cluster \SPT\ (also known
as PLCKG266.6$-$27.3) in 2023 September (JWST-GO-4212; PI Bradley). The
cluster has a redshift of $\zspec = 0.972$, and has a very large mass of
$M_{500} = 7.1 \times 10^{14} \Msun h_{70}^{-1}$ \citep{Williamson2011}
for its distance.

The NIRCam observations include four short-wavelength (SW) filters
(F090W, F115W, F150W, and F200W) and four long-wavelength (LW) filters
(F277W, F356W, F410M, and F444W) spanning $0.8 - 5.0~\micron$ with
2920.4~s of exposure time in each filter. Each exposure uses the MEDIUM8
readout pattern with seven groups/integration and one integration.
Four dithers were obtained with the INTRAMODULEBOX dither pattern,
designed to fill the 5\arcsec\ gaps in the NIRCam SW detectors and
maximize the area with full exposure time. The dithers mitigate the
effects of bad pixels, image artifacts, and flat-field uncertainties.
They also improve the spatial resolution of the resampled/drizzled
images. The NIRCam imaging includes two $2\farcm3 \times 2\farcm3$
fields separated by 40\farcs5, covering 10.2 arcmin$^{2}$ in total.
The \SPT\ cluster was centered on NIRCam module B, while NIRCam module
A obtained observations on a nearby field centered $\sim$ 2\farcm9
north-northwest of the cluster center. The JWST observations are shown
in Figures~\ref{fig:cluster} and \ref{fig:parallel} and summarized in
Table~\ref{tbl:obs}.

\begin{deluxetable*}{llcccc}
\tablecaption{HST and JWST Exposure Times and Depths
\label{tbl:obs}}
\tablewidth{\columnwidth}
\tablehead{
\colhead{} &
\colhead{} &
\colhead{} &
\colhead{} &
\colhead{Cluster} &
\colhead{Parallel}
\\[-6pt]
\colhead{} &
\colhead{} &
\colhead{Wavelength} &
\colhead{Exposure Time} &
\colhead{$m_{\mathrm{lim}}$\tnm{a}} &
\colhead{$m_{\mathrm{lim}}$\tnm{a}}
\\[-6pt]
\colhead{Camera} &
\colhead{Filter} &
\colhead{(\micron)} &
\colhead{(s)} &
\colhead{(AB)} &
\colhead{(AB)}
}
\startdata
HST ACS/WFC & F435W & 0.37--0.47 & ~\,2249        & 27.8 & 28.1\tnm{b} \\
HST ACS/WFC & F606W & 0.47--0.7  & ~\,8880\tnm{c} & 28.7 & 28.7 \\
HST ACS/WFC & F814W & 0.7--0.95  &  12,720\tnm{d} & 28.7 & 28.3 \\
HST WFC3/IR & F105W & 0.9--1.2   & ~\,4166\tnm{e} & 28.6 & \nodata \\
HST WFC3/IR & F125W & 1.1--1.4   & ~\,3464\tnm{e} & 28.6 & \nodata \\
HST WFC3/IR & F140W & 1.2--1.6   & ~\,5874\tnm{e} & 29.1 & \nodata \\
HST WFC3/IR & F160W & 1.4--1.7   & ~\,7374\tnm{e} & 28.8 & \nodata \\
JWST NIRCam & F090W & 0.8--1.0   & ~\,2920        & 28.9 & 29.0 \\
JWST NIRCam & F115W & 1.0--1.3   & ~\,2920        & 28.9 & 29.0 \\
JWST NIRCam & F150W & 1.3--1.7   & ~\,2920        & 29.2 & 29.2 \\
JWST NIRCam & F200W & 1.7--2.2   & ~\,2920        & 29.4 & 29.4 \\
JWST NIRCam & F277W & 2.4--3.1   & ~\,2920        & 29.7 & 29.7 \\
JWST NIRCam & F356W & 3.1--4.0   & ~\,2920        & 29.7 & 29.7 \\
JWST NIRCam & F410M & 3.8--4.3   & ~\,2920        & 29.0 & 29.0 \\
JWST NIRCam & F444W & 3.8--5.0   & ~\,2920        & 29.3 & 29.3 \\
\enddata
\vspace{1em}
Notes.
\tablenotetext{a}{5$\sigma$ limiting AB magnitude in a $r=0\farcs1$ circular aperture measured on the background-subtracted data.}
\tablenotetext{b}{Only a small corner of the F435W image covers the parallel field.}
\tablenotetext{c}{Total exposure time includes an overlapping $2 \times 2$ mosaic centered on the cluster (GO~12477; 1920~s each) and a pointing centered on ESO~121$-$20 in the parallel field (GO~9771; 1200~s). The quoted depths represent an average over the mosaic.}
\tablenotetext{d}{Total exposure time includes an overlapping $2 \times 2$ mosaic centered on the cluster (GO~12757; 2476~s each), a single pointing centered on the cluster (GO~12477; 1916~s), and a pointing centered on ESO~121$-$20 in the parallel field (GO~9771; 900~s). The quoted depths represent an average over the mosaic.}
\tablenotetext{e}{Total exposure time after removing MULTIACCUM reads affected by scattered earthshine.}
\end{deluxetable*}

\subsection{HST Data}

We supplement the JWST NIRCam observations with archival HST
optical and near-infrared imaging of \SPT. Both the South Pole
Telescope Survey \citep[GO~12477;][]{Williamson2011} and the Planck
Collaboration \citep[GO~12757;][]{Planck2011} obtained HST imaging
of \SPT\ with Advanced Camera of Surveys/Wide Field Camera (ACS/WFC)
F606W and F814W in 2013 January. The RELICS HST Treasury program
\citep[GO~14096;][]{Coe2019} obtained HST imaging of \SPT\ with ACS/WFC
F435W (one orbit) and WFC3/IR F105W, F125W, F140W, and F160W (two orbits
total) in 2017. Additional HST WFC3/IR imaging was obtained in F105W
(one orbit), F125W (one orbit), F140W (two orbits), and F160W (two
orbits) by the RELICS team in 2020 (GO~15920; PI Salmon).

The HST observations are summarized in Table~\ref{tbl:obs}. In total,
the JWST and HST observations of \SPT\ include imaging in 15 filters
spanning $0.4 - 5.0~\micron$.


\begin{figure}[t!]
\begin{center}
\includegraphics[width=0.5\textwidth]{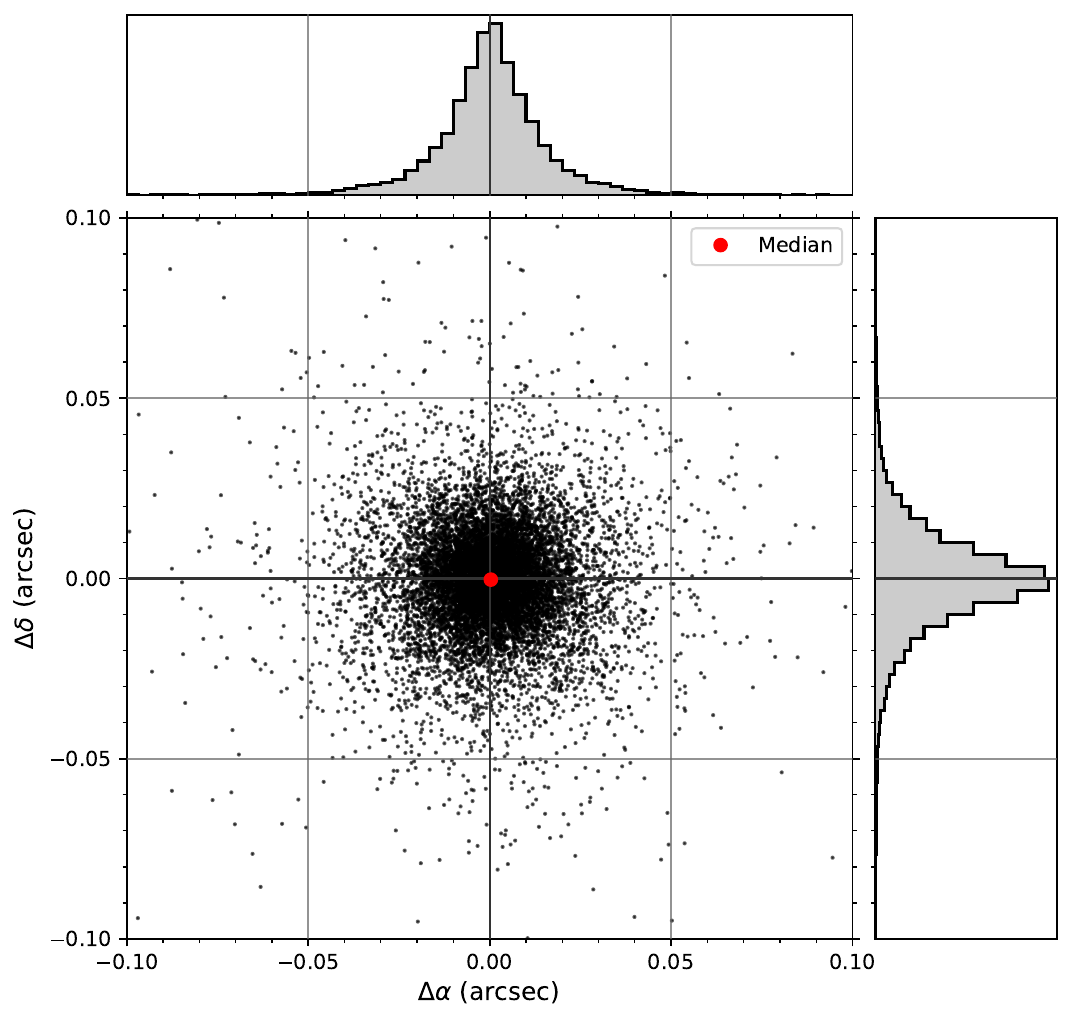}
\end{center}
\caption{
Diagnostic plot of the astrometric alignment between the JWST
NIRCam F200W (SW) and F277W (LW) images. The black points in the
scatter plot show the source position offsets in the F200W and F277W
40~mas~\perpixel\ mosaic images. The red circle denotes the median
astrometric offset, which is consistent with zero. The histograms
show the distribution of offsets in R.A. ($\Delta \alpha$) and decl.
($\Delta \delta$). We find 99\% of all sources detected in these images
have positional offsets smaller than 10~mas, with an rms of 22~mas.
Restricting the sample to bright (SNR $> 10$), compact (FWHM $<$
0\farcs1) sources yields an rms of just 8~mas, demonstrating excellent
astrometric alignment.
\label{fig:astrometry}
}
\end{figure}

\section{Methods} \label{sec:methods}

\subsection{Data Reduction}  \label{sec:reduction}

We reduced the pipeline-calibrated HST data and the JWST level-2
imaging products using the \grizli\ (version 1.9.5) reduction package
\citep{Grizli}. The JWST data were processed with version 1.11.4 of the
calibration pipeline with CRDS context \texttt{jwst\_1123.pmap}, which
includes photometric calibrations based on in-flight data.

The \grizli\ pipeline reprocesses the HST WFC3/IR data to correct
exposures affected by time-variable sky backgrounds caused by scattered
earthshine. For the NIRCam data, the \grizli\ pipeline applies a
correction to reduce the effect of $1/f$ noise, masks ``snowballs''
caused by large cosmic-ray impacts, and subtracts templates to remove
the ``wisp'' stray-light features from the NIRCam SW detectors.

Astrometric alignment of the HST and JWST images was performed using
the \grizli\ package, which computes windowed centroid coordinates
of sources with the \SEP\ package \citep{Barbary2016} to achieve
precise alignment. The HST and JWST data were aligned to a common World
Coordinate System registered in the GAIA DR3 catalog \citep{Gaia_EDR3}.

We combined and resampled the fully calibrated images in each filter
to a common pixel grid using \astrodrizzle\ \citep{MultiDrizzle,
DrizzlePac}. The HST and JWST NIRCam LW filter images were drizzled
to a grid with 0$\farcs$04 \perpixel, while the JWST NIRCam SW filter
images, with their smaller native pixel scale, were drizzled to a grid
with 0$\farcs$02 \perpixel. In subsequent steps, we rebin the NIRCam SW
images to a pixel scale of 0$\farcs$04 \perpixel\ to place the images
for all 15 filters on the same pixel-registered grid.

In Figure~\ref{fig:astrometry}, we show a diagnostic plot of the
astrometric alignment between the JWST NIRCam F200W (SW) and F277W (LW)
40~mas~\perpixel\ mosaic images. We find 99\% of all sources detected
in these images have separations less than 10~mas, with an rms 22~mas.
Selecting bright (SNR $> 10$), compact (FWHM $<$ 0\farcs1) sources,
yields an rms of only 8~mas, indicating excellent astrometric alignment.
This is consistent with the expected astrometric alignment accuracy of
JWST NIRCam images \citep[e.g.,][]{Bagley2023}.


Our NIRCam SW images show the presence of stronger-than-usual wisp
features that were not sufficiently removed by the wisp-template
subtraction. Wisps are caused by off-axis light from bright stars
reflecting off the top secondary mirror strut of JWST and entering
the Aft-Optics-System mask. The wisp geometry and intensity can vary
significantly depending on the exact telescope pointing. Wisps are
present only in NIRCam SW images and are most prominent in the A3, A4,
B3, and B4 detectors \citep{Rigby2023}.

To remove the residual wisps from the NIRCam SW data, we use an
iterative two-dimensional (2D) background procedure using the
\photutils\ Background2D and SourceFinder classes. This approach
closely follows the method used by the CEERS team, as described in
\citet{Bagley2023}, who also employ an iterative masking process in
combination with the Background2D class. Multiple iterations are
performed to refine the source mask, which is then used to exclude
sources during background estimation in small spatial boxes.

We start by computing a 2D background from the 20~mas~\perpixel\ SW
drizzled images using the Background2D class with large box sizes ($100
\times 100$ pixels), median filtering over $3 \times 3$ boxes, and a
sigma-clipping threshold of 3$\sigma$. We then use the SourceFinder
class (with deblending turned off) to detect sources in the image
larger than 50 pixels with a 2D threshold image calculated as the 2D
background image plus a multiple of the background rms image. We then
create a source mask from the segmentation image and dilate it using
a circular footprint with a radius of 31 pixels. The dilated source
mask is then input into the Background2D class with a smaller box size
to compute a new background-subtracted image, from which a new source
mask is created. This iterative process was repeated 3 times with box
sizes of 100, 50, and 25 pixels, detection thresholds of 3.0$\sigma$,
3.0$\sigma$, and 1.8$\sigma$, detection pixel sizes of 50, 9, and 9
pixels, and dilation sizes of 31, 21, and 15 pixels, respectively.
The final source mask was then used to compute and subtract the final
background image using the Background2D class with a box size of $10
\times 10$ and median filtered over $5 \times 5$ boxes.

\begin{figure*}[t!]
\plotone{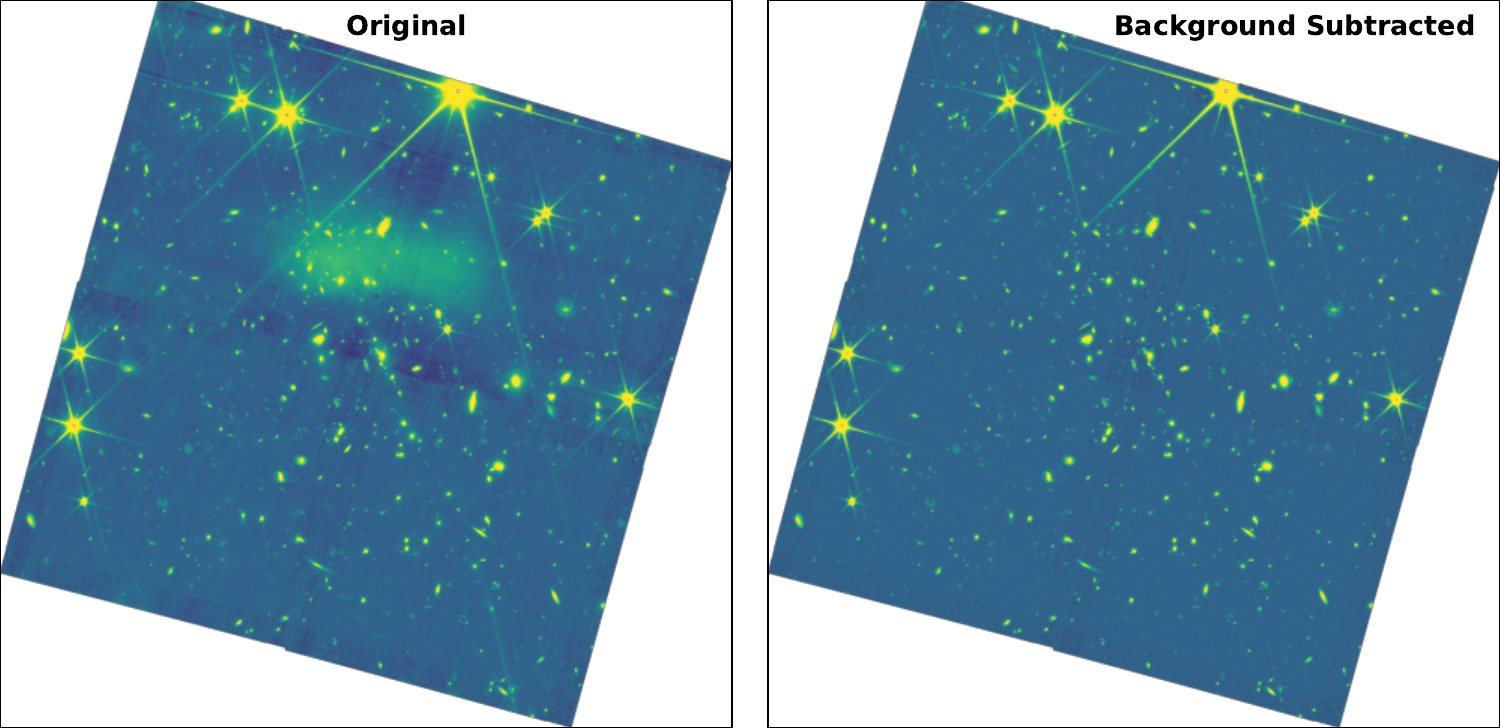}
\caption{
JWST F090W image of the \SPT\ cluster field demonstrating the background
subtraction. The image on the left is the original mosaic image before
background subtraction. Note the presence of a strong wisp residual
as described in Section~\ref{sec:reduction}. This feature and other
nonuniform background features are removed in the background-subtracted
image on the right. A residual wisp feature is present in all of the
original NIRCam SW images, but it is most prominent in the F090W data.
\label{fig:bkgsub}
}
\end{figure*}

Residual wisps were present in all of our NIRCam SW images. It is
strongest in F090W, but becomes less prominent as the wavelength
increases for the other SW filters. We show an example of the background
removal in Figure~\ref{fig:bkgsub} using our worst-case F090W image. The
residual wisp feature is clearly visible in the original image on the
left. The background subtraction procedure performs well in eliminating
residual wisps and other spatially varying background components. While
it also subtracts intracluster light, this does not impact our analysis
of the \arc\ or its counterimage candidate.

\subsection{Photometric Catalog of sources in the \SPT\ region} \label{sec:photcat}

Sources were identified in a detection image composed of an inverse
variance-weighted sum of the F277W, F356W, and F444W NIRCam LW images
using \photutils\ \citep{photutils111, photutils220} image-segmentation
tools. Before performing photometry, we rebinned the NIRCam SW images to
a pixel scale of 0$\farcs$04~\perpixel\ to place the images for all 15
filters on the same pixel-registered grid.

Our photometric catalog was created using the same procedure as
described in \cite{Bradley2023}. Photometry was measured in all bands
using \photutils\ SourceCatalog with the segmentation image and catalog
produced from the detection image. Source colors were measured in
elliptical Kron apertures with a scale factor of 1.5 to accurately
recover the colors of distant galaxies. Each source has a different
elliptical-aperture size and orientation based on the central moments
of its flux distribution in the detection image. We derived aperture
corrections by computing the ratio of the flux in a larger Kron aperture
(with a Kron scale factor of 2.5) to that in the smaller aperture
for each source, as measured in the detection image. We applied this
aperture correction to the fluxes and uncertainties for all filters
to compute total Kron fluxes. Isophotal fluxes were also calculated
by summing the fluxes within the source segments defined by the
segmentation image. Additional correction factors are not applied to the
isophotal fluxes.

\subsection{Photometric Redshifts} \label{sec:photoz}

We measure photometric redshifts using \eazypy\ \citep{Brammer2008}
for all sources in our catalog using the photometry measured in the
elliptical Kron apertures. \eazypy\ fits the observed photometry of
each galaxy with a nonnegative linear combination of templates to
derive a probability distribution function of the redshift. We use a
template set composed of the 12 ``tweak\_fsps\_QSF\_12\_v3'' templates
derived from the Flexible Stellar Population Synthesis (FSPS) library
\citep{Conroy2009, Conroy2010a, Conroy2010b}, which includes a range
of galaxy types (e.g., star-forming, quiescent, dusty) and realistic
star formation histories (SFHs) (e.g., bursts, slowly rising, slowly
falling). We also include six additional templates (sets 1 and 4) from
\cite{Larson2023} that are based on a combination of Binary Population
and Spectral Synthesis (BPASS) and CLOUDY models. These templates have
bluer colors than the fiducial FSPS templates and match the predicted
rest-UV colors of simulated galaxies at $z > 8$. The inclusion of these
additional templates provides improvements in the photometric redshift
accuracy for bluer galaxies at $z > 8$ \citep{Larson2023}.

We assume a flat luminosity prior, similar to recent JWST high-redshift
studies \citep[e.g.,][]{Finkelstein2022, Finkelstein2023a, Adams2024},
to prevent bias against selecting bright high-redshift galaxies,
whose luminosity function is poorly known. We apply an error floor of
5\% to the flux uncertainties to account for photometric calibration
uncertainties. We allow the redshifts to span from $0.1 < z < 20$,
in steps of 0.01. We also perform a second run of \eazypy\ with the
redshift range restricted to $z < 7$ to compare the fiducial results
with the best-fit low-redshift solutions.

To check the accuracy of our photometric redshifts, we compared
them with literature spectroscopic redshifts for sources in the
\SPT\ cluster. For source 1.1, our \zphot\ = $1.4_{-0.1}^{+0.1}$
(95\%) is in excellent agreement with its spectroscopic value of
\zspec\ = 1.358 \citep{Paterno-Mahler2018}. The system comprising
sources 3.1 and 3.2 has a spectroscopic redshift of \zspec\ = 4.013
\citep{Paterno-Mahler2018}. Our photometric redshifts, \zphot\ =
$4.1_{-0.1}^{+0.2}$ (95\%) for 3.1 and $4.3_{-0.2}^{+0.1}$ (95\%) for
3.2, are consistent with this value. While the \zspec\ falls within the
uncertainty for source 3.1, it lies just outside the 95\% confidence
interval for source 3.2.

A recent study comparing photometric and spectroscopic redshifts for
43 $z = 7 - 13$ galaxies measured with \eazypy\ from the JWST JADES
\citep{Rieke2023} and CEERS \citep{Finkelstein2023a} surveys found
that the photometric redshifts show excellent agreement with the
spectroscopic redshifts \citep{Duan2024}. Overall, 4.6\% (2 of 43)
of the sources qualify as outliers, defined as having $\zphot > 1.15
(\zspec + 1)$ or $\zphot < 0.85 (\zspec + 1)$ \citep{Duan2024}. Both of
the outlier sources were at $7 < \zphot < 8$. None of the sources at $z
> 8$ were classified as outliers.

JWST imaging studies have been effective in selecting high-redshift
galaxies using photometric redshifts. While recent JWST spectroscopic
results show a high confirmation rate for these candidates, several
independent analyses report that the photometric redshifts of galaxies
at $z > 7$ are systematically overestimated by $\Delta z \sim 0.2 - 0.3$
or more relative to spectroscopic measurements, with the largest offsets
occurring at $z > 9$ \citep[e.g.,][]{ArrabalHaro2023, Fujimoto2023,
Finkelstein2024, Hainline2024, Helton2024, Willott2024}. For example, at
$\zphot > 9.7$ \cite{Finkelstein2024} found large offsets of $\Delta z =
\zphot - \zspec$ of 0.5 (CEERS-11384) and 1.2 (CEERS-19996).

This systematic offset is likely caused by increased \Lya\ damping
wing absorption, which produces a smoother spectral break at
rest-frame \Lya\ (1216~\AA) than the sharp break commonly employed
\citep[e.g.,][]{Madau1995, Inoue2014} in photometric redshift and
spectral energy distribution (SED)-fitting codes. This absorption
is attributed to either the intergalactic medium during the Epoch
of Reionization (EoR) or nearby damped Lyman-alpha systems along
the line of sight \citep[e.g.,][L. Christensen et al. 2025, in
preparation]{Curtis-Lake2023, Hainline2024, Heintz2024, Hsiao2024,
Umeda2024, Asada2025, Messa2025}.

\section{Cluster Lens Models} \label{sec:lens_models}

We generated four independent lensing models for the foreground
strong-lensing galaxy cluster \SPT\ to estimate source magnifications
using \lenstool, \glafic, and \wslap. These new models are
improvements over the previous \SPT\ lens models presented in
\cite{Paterno-Mahler2018} and \cite{Salmon2018}, which were based
only on HST imaging data. The different models not only span a range
of modeling algorithms. Their independent construction also gives
us leverage over systematic uncertainties that are due to modeling
choices. The models differed in how they considered the observational
constraints (e.g., whether to include candidate lensed images;
photometric redshifts; including clumps within images of lensed galaxies
as individual constraints; astrometric uncertainty), and choices for the
parameterization of the lens plane (number of halos, substructure, and
interloping masses).

\subsection{\lenstool\ Models}

We generated two different lensing models using \lenstool\
\citep{Jullo2007}, which we refer to as \lta\ and \ltb. \lenstool\
employs a parametric approach and Markov Chain Monte Carlo (MCMC)
sampling of the parameter space to identify the best-fit model and
associated uncertainties.

For the \lta\ model, the cluster lens is represented by a combination
of three main halos with contributions from cluster member galaxies,
all parameterized as pseudo-isothermal ellipsoidal mass distributions
\citep{Limousin2005}. Galaxy-scale halo parameters are determined
using scaling relations \citep{Jullo2007}. The parameters of the two
cluster-scale halos are allowed to vary, except for the cut radius,
which is larger than the strong-lensing region and thus cannot be
constrained by strong-lensing evidence. The positional parameters
of the brightest cluster galaxy (BCG) halo are fixed to observed
values, while its slope parameters are allowed to vary. The model
incorporates constraints from the positions of 43 multiple images
belonging to 14 clumps from nine distinct source galaxies. Redshifts
of two spectroscopically confirmed sources at $z=1.358$ and $z=4.013$
\citep{Paterno-Mahler2018} and the \arc\ at $z=10.2$ are used as
constraints (the A, B, and C star clusters from \cite{Adamo2024}; see
Figure~\ref{fig:arc}). The redshifts of arcs without spectroscopic
redshifts are treated as free parameters with very broad priors. While
the model predicts a counterimage at (R.A., decl.)=($93.9490607$,
$-57.7701814$), a potential candidate (see Figure~\ref{fig:cluster})
observed near this location ($\sim 1\farcs8$) was not used as a
constraint.

External shear is not required. However, one of the cluster-scale
halos is generally aligned with the galaxy distribution of the
foreground group at $z=0.42$ \citep{Teja2023}, possibly accounting for
contributions from this structure. All observed lensed features are well
reproduced by this model. The image plane rms of the best-fit \lta\
model is 0\farcs36. The \lta\ model is used as the reference model for
the analysis presented in this paper.

The \ltb\ cluster lens model uses a different set of input assumptions,
including a different position of the mass distribution of the
lens. This model uses 43 multiple images from 11 unique sources as
constraints. A secondary cluster-scale halo is placed around the
location of dusty galaxies nearly 50\arcsec\ north of the BCG. The
position of this halo is allowed to move within a 20\arcsec\ box around
this position. The image plane rms of the best-fit \ltb\ model is
0\farcs68.

\subsection{\glafic\ Model}

We construct another mass model using \glafic\ \citep{Oguri2010,
Oguri2021}. We follow the methodology described in \citet{Kawamata2016}
to determine a set of lens mass components used for mass modeling.
Our mass model consists of three elliptical Navarro–Frenk–White
(NFW) halos \citep{Navarro1997}, external shear, and cluster member
galaxies modeled by pseudo-Jaffe ellipsoids. We fix the positions
of two NFW halos at the locations of bright cluster member galaxies
at (R.A., decl.) = ($93.9663980$, $-57.7791580$) and ($93.9705078$,
$-57.7753866$), while the center of the third NFW mass halo is left as a
free parameter. For all three elliptical NFW components, we leave their
masses, ellipticities, position angles, and concentration parameters
as free parameters. In order to reduce the number of parameters,
we adopt the standard scaling relations between the luminosities,
velocity dispersions, and truncation radii of cluster member galaxies
\citep[see][for more details]{Kawamata2016}. The observational
constraints consist of the positions of 44 multiple images from 15
background sources. Spectroscopic redshifts are available for five of
the 15 background sources \citep{Paterno-Mahler2018}. We fix the source
redshift of the \arc\ to $z=10.2$. In addition, we include constraints
on redshifts of six background sources based on their photometric
redshifts, with a conservative error on the photometric redshifts of
$\sigma_z=0.5$ assuming the Gaussian distribution. In order to better
reproduce the shape of the \arc, we also include the positions of the
star cluster systems A.1/A.2 and B.1/B.2 (see Figure~\ref{fig:arc}) in
the \arc\ as constraints. We assume a positional error of 0\farcs4,
except for the positions of A.1/A.2 and B.1/B.2, for which we assume a
smaller positional error of 0\farcs04.

The best-fitting model has $\chi^2=52.7$ for 37 dof, and reproduces
all multiple image positions well, with an rms of image positions of
0\farcs41.

\subsection{\wslap\ Model}

The \wslap\ model \citep{Diego2005, Diego2007} is a hybrid model
combining a large-scale component for the mass and a small-scale
component. The large-scale distribution of the mass is described by
a predetermined grid (uniform or multiresolution) of 2D Gaussian
functions. The small-scale component follows directly the light
distribution of member galaxies. The \wslap\ lens model offers an
alternative to parametric models and is free of assumptions made about
the distribution of dark matter.

For the small scale, we select the most luminous elliptical member
galaxies and assume their mass follows the distribution of light in the
F356W filter. All galaxies are distributed in three groups, each group
with a fixed light-to-mass ratio that is optimized by \wslap. The first
group contains the central BCG. The second group contains bright member
galaxies at $z=0.972$ ($N\approx 70$). Finally, the third group contains
two foreground galaxies at $z=0.42$ that are near other arcs in the
background. This third group has a negligible impact on the \arc, but it
is added to the lens model to increase its precision in other portions
of the image.

The smooth component is composed of either a $20 \times 20$ regular
grid of 2D Gaussians or a 158 grid of 2D Gaussians. The second grid is
derived from a solution obtained with the regular grid and has increased
resolution in the central region of the cluster. In both cases, the
amplitudes of the Gaussians are optimized by \wslap. We use a set of
14 multiple images as constraints from 11 individual galaxies. As in
other lens models in this work, we use multiply lensed knots in several
galaxies as additional constraints. The rms of the model is $\approx
1\arcsec$ (image plane), although for two systems, we find an rms of
$\approx 2\arcsec$, indicating a possible tension in that portion of the
lens plane. These two systems are far from the $z \sim 10.2$ \arc.

\section{Results and Discussion} \label{sec:results}

This section focuses on the analysis of the entire galaxy appearing as
the \arc. We first present its photometric properties and discuss its
photometric redshift. We describe the estimated magnifications of the
arc and the detection of a candidate counterimage. Finally, we perform
SED fitting to the multiband photometry of the \arc\ and discuss the
intrinsic physical properties of the galaxy.

\begin{figure*}[t!]
\begin{center}
\includegraphics[width=0.98\textwidth]{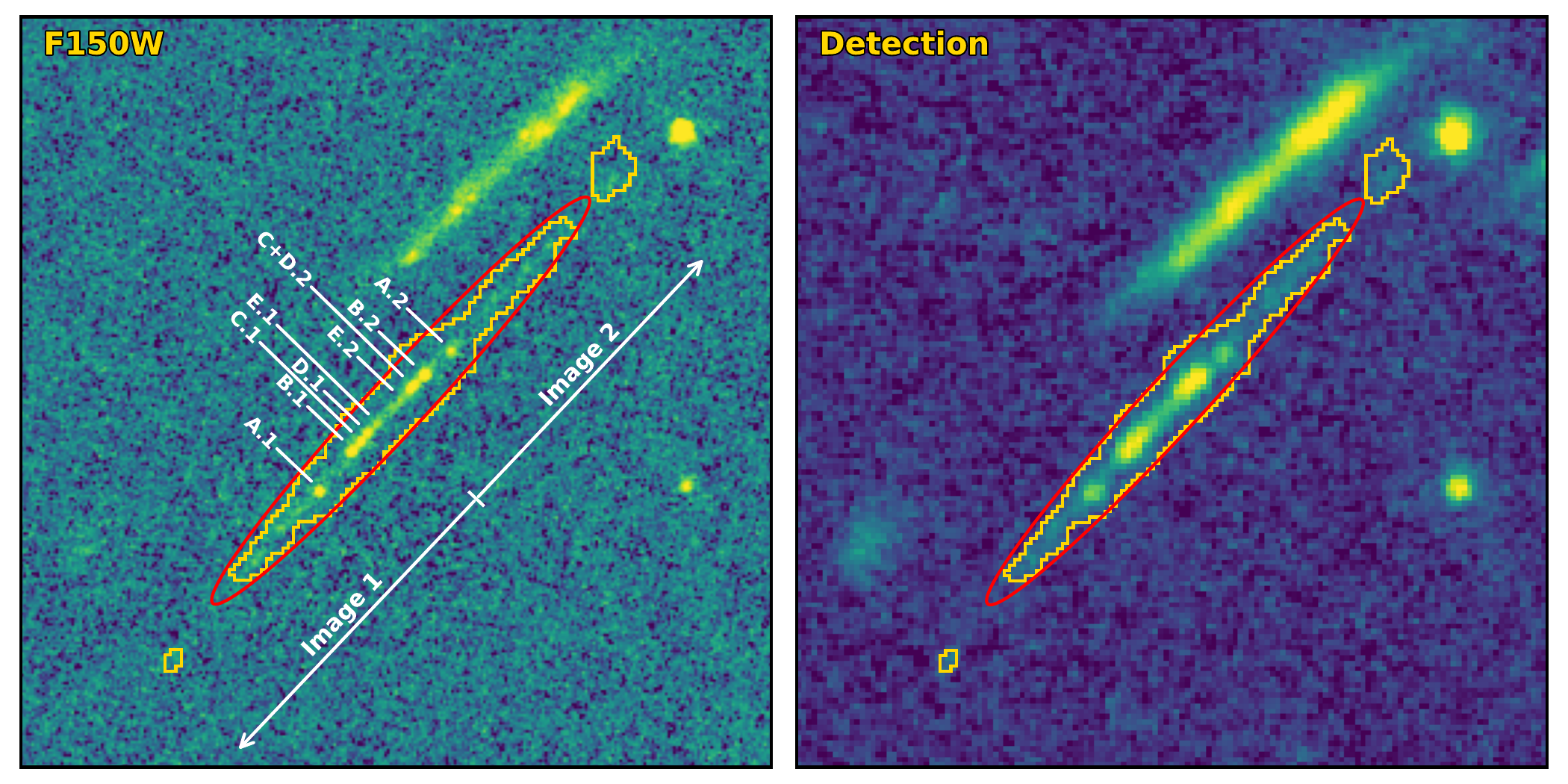}
\end{center}
\caption{
Cutout images (4\farcs8 $\times$ 4\farcs8) of the \arc\ in the NIRCam
F150W and detection (F277W + F356W + F444W) image. The source segments,
outlined in gold, were defined from the detection image, which has a
very broad PSF. The Kron ellipse, shown in red, was defined from the
central moments of the flux distribution within the source segments.
In the F150W image, we also label the two mirrored images of the arc,
Images 1 and 2, as well as the star clusters A through E analyzed in
detail by \cite{Adamo2024}.
\label{fig:arc}
}
\end{figure*}

\begin{figure*}[t!]
\begin{center}
\includegraphics[width=0.98\textwidth]{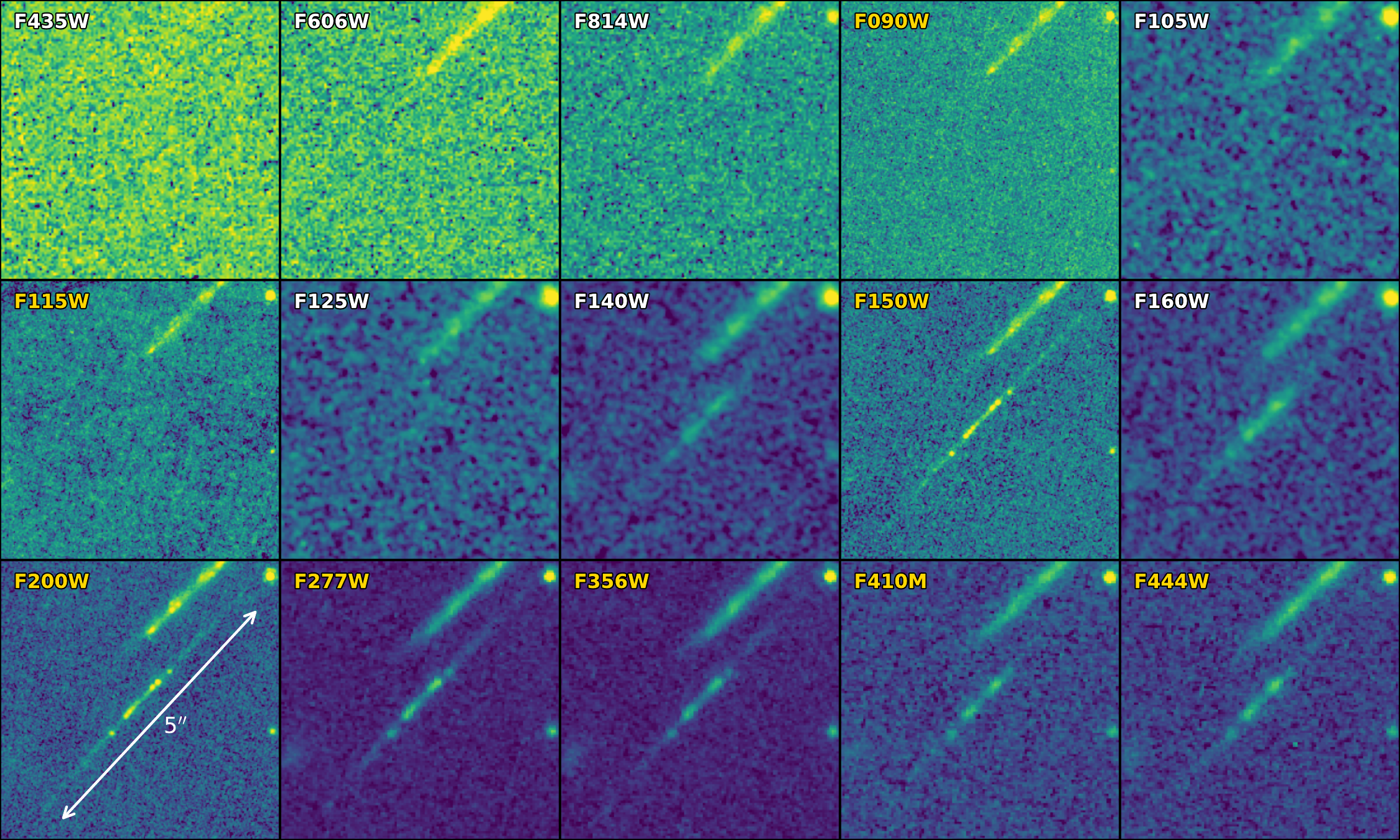}
\end{center}
\caption{
Cutout images of the \arc\ in the observed HST and JWST filters. The
cutout field of view is 4\farcs8 $\times$ 4\farcs8, and the images are
shown with north up and east left. Filters are labeled in the upper-left
corner of each panel with white text for HST images and gold for JWST
images. The \arc\ extends to 5\arcsec\ long in the JWST NIRCam images
and is centered in the cutout images. The galaxy is undetected in F115W
and all bluer filters. It is weakly detected in F125W and clearly
detected in F140W and all redder filters. The JWST NIRCam F150W image
(PSF FWHM 0\farcs05) provides the best resolution of the \arc, revealing
star clusters down to 1~pc in intrinsic size \citep{Adamo2024}. The
nearby arc to the north has a photometric redshift of $z \sim 2.6$.
\label{fig:cutouts}
}
\end{figure*}

\subsection{The \arc\ (\sptarc)}

In Figure~\ref{fig:cutouts}, we show cutout images ($4\farcs8 \times
4\farcs8$) of the \arc\ in the observed HST and JWST filters. The galaxy
is weakly detected in the WFC3/IR F125W image and clearly detected in
F140W and all redder filters. The \arc\ is undetected in the JWST NIRCam
F115W image and all bluer filters. While \cite{Salmon2018} reported the
length of the \arc\ to be 2\farcs5 from the HST WFC3/IR F160W data, the
JWST data reveal faint emission that extends to 5\farcs0 long in the
NIRCam images.

The JWST NIRCam F150W image provides the best resolution of the \arc,
revealing at least nine observed bright knots along the arc (see
Figures~\ref{fig:cutouts} and \ref{fig:arc}). Overall, there is a
distinctive symmetry of the knot locations from the center of the
extended arc, suggesting that the observed arc represents two mirror
images of the lensed galaxy. This is confirmed by our cluster lens
models (see Section~\ref{sec:lens_models}), all of which predict the
$z=10.2$ critical curve crossing the arc (see Figure~\ref{fig:cluster}
inset).

The bright knots in the \arc\ have been identified as individual star
clusters, resolved down to 1~pc in intrinsic size in the JWST NIRCam
F150W image \citep{Adamo2024}. This is the first such detection of star
clusters at $z > 10$, only 460~Myr after the Big Bang, made possible by
the combination of JWST's high sensitivity and spatial resolution and
the gravitational lensing magnification of the \arc. The clusters have
very large stellar surface densities of $\sim 10^5~\Msun$~pc$^{-2}$
and are consistent with gravitationally bound YSCs \citep{Adamo2024}.
These clusters could be the progenitors of metal-poor globular clusters
at $z=0$. The possibility that some progenitors of today's globular
clusters have formed in low-mass galaxies at $z>7$ is particularly
interesting also because it is somewhat at odds with the predictions of
several theoretical models \citep[e.g.,][]{Reina-Campos2019}, with a few
exceptions \citep{Katz2013, Katz2014}. Importantly, the high stellar
densities within these systems pave the way to massive stars and stellar
black hole runaway collisions, which might lead to the formation of
intermediate-mass black hole seeds \citep{gieles2018, antonini2019}.

\begin{deluxetable*}{ccccccccc}
\tablecolumns{9}
\tablecaption{Measured Photometry of \sptarc\ with HST}
\label{tbl:hstphot}
\tablehead{\colhead{Source\tnm{a}} & \colhead{\zphot\tnm{b}} & \colhead{F435W} & \colhead{F606W} & \colhead{F814W} & \colhead{F105W} & \colhead{F125W} & \colhead{F140W} & \colhead{F160W}\\[-0.5em] \colhead{} & \colhead{ } & \colhead{$\mathrm{(nJy)}$} & \colhead{$\mathrm{(nJy)}$} & \colhead{$\mathrm{(nJy)}$} & \colhead{$\mathrm{(nJy)}$} & \colhead{$\mathrm{(nJy)}$} & \colhead{$\mathrm{(nJy)}$} & \colhead{$\mathrm{(nJy)}$}}
\startdata
Cosmic Gems (Kron) & $10.19_{-0.16}^{+0.16}$ & $5.1 \pm 32.7$ & $-17.4 \pm 19.9$ & $-8.0 \pm 18.0$ & $38.1 \pm 24.1$ & $123.8 \pm 24.8$ & $292.4 \pm 16.1$ & $543.0 \pm 21.2$ \\
Cosmic Gems (Isophotal) & $10.18_{-0.16}^{+0.16}$ & $12.0 \pm 26.0$ & $-0.6 \pm 15.9$ & $-3.2 \pm 14.3$ & $33.0 \pm 19.2$ & $96.1 \pm 19.7$ & $248.1 \pm 12.9$ & $447.9 \pm 16.9$ \\
Counterimage\tnm{c} & $10.8_{-1.4}^{+0.6}$ & $2.8 \pm 7.2$ & $-3.9 \pm 4.2$ & $-2.5 \pm 4.7$ & $1.5 \pm 6.2$ & $-0.6 \pm 5.7$ & $7.8 \pm 3.5$ & $9.7 \pm 5.3$ \\
\hline
\sidehead{PSF-matched Photometry}
\hline
Cosmic Gems (Kron) & $10.23_{-0.18}^{+0.40}$ & $0.0 \pm 38.5$ & $-19.6 \pm 23.4$ & $-8.4 \pm 21.1$ & $5.9 \pm 28.2$ & $97.0 \pm 29.1$ & $281.8 \pm 18.8$ & $546.4 \pm 27.3$ \\
Cosmic Gems (Isophotal) & $10.17_{-0.19}^{+0.37}$ & $-8.1 \pm 34.4$ & $-18.9 \pm 20.9$ & $3.2 \pm 18.8$ & $17.6 \pm 25.2$ & $89.6 \pm 26.0$ & $265.6 \pm 16.8$ & $488.0 \pm 24.4$ \\
\enddata
\tablecomments{Observed fluxes, uncorrected for magnification. $\mAB = 31.4 - 2.5 \log(f_{\nu} / {\rm nJy}).$}
\tablenotetext{a}{Figure~\ref{fig:arc} shows the Kron aperture and isophotal segments overplotted on the arc for the non-PSF-matched photometry.}
\tablenotetext{b}{Photometric redshift measured with \eazypy. Errors are 95\% confidence intervals.}
\tablenotetext{c}{Photometry measured in the Kron aperture described in Section~\ref{sec:photcat}.}
\end{deluxetable*}

\begin{deluxetable*}{ccccccccc}
\tablecolumns{9}
\tablecaption{Measured Photometry of \sptarc\ with JWST/NIRCam}
\label{tbl:jwstphot}
\tablehead{\colhead{Source\tnm{a}} & \colhead{F090W} & \colhead{F115W} & \colhead{F150W} & \colhead{F200W} & \colhead{F277W} & \colhead{F356W} & \colhead{F410M} & \colhead{F444W}\\[-0.5em] \colhead{ } & \colhead{$\mathrm{(nJy)}$} & \colhead{$\mathrm{(nJy)}$} & \colhead{$\mathrm{(nJy)}$} & \colhead{$\mathrm{(nJy)}$} & \colhead{$\mathrm{(nJy)}$} & \colhead{$\mathrm{(nJy)}$} & \colhead{$\mathrm{(nJy)}$} & \colhead{$\mathrm{(nJy)}$}}
\startdata
Cosmic Gems (Kron) & $1.9 \pm 14.0$ & $-4.8 \pm 12.9$ & $508.5 \pm 11.6$ & $574.0 \pm 9.1$ & $457.3 \pm 8.0$ & $368.5 \pm 7.9$ & $357.1 \pm 15.3$ & $402.2 \pm 12.5$ \\
Cosmic Gems (Isophotal) & $-7.5 \pm 11.1$ & $4.2 \pm 10.3$ & $442.3 \pm 9.2$ & $491.4 \pm 7.4$ & $396.3 \pm 6.5$ & $321.6 \pm 6.3$ & $302.2 \pm 12.3$ & $350.3 \pm 10.0$ \\
Counterimage\tnm{b} & $1.9 \pm 3.1$ & $0.2 \pm 3.0$ & $13.5 \pm 2.5$ & $15.3 \pm 2.0$ & $11.0 \pm 1.5$ & $8.4 \pm 1.5$ & $8.8 \pm 3.0$ & $6.5 \pm 2.4$ \\
\hline
\sidehead{PSF-matched Photometry}
\hline
Cosmic Gems (Kron) & $-3.6 \pm 16.4$ & $-6.0 \pm 15.2$ & $452.3 \pm 22.6$ & $534.4 \pm 26.7$ & $436.5 \pm 21.8$ & $365.2 \pm 18.3$ & $329.6 \pm 17.9$ & $401.4 \pm 20.1$ \\
Cosmic Gems (Isophotal) & $-3.1 \pm 14.7$ & $-1.4 \pm 13.6$ & $407.3 \pm 20.4$ & $484.6 \pm 24.2$ & $394.0 \pm 19.7$ & $334.6 \pm 16.7$ & $301.0 \pm 15.9$ & $366.2 \pm 18.3$ \\
\enddata
\tablecomments{Observed fluxes, uncorrected for magnification. $\mAB = 31.4 - 2.5 \log(f_{\nu} / {\rm nJy}).$}
\tablenotetext{a}{Figure~\ref{fig:arc} shows the Kron aperture and isophotal segments overplotted on the arc for the non-PSF-matched photometry.}
\tablenotetext{b}{Photometry measured in the Kron aperture described in Section~\ref{sec:photcat}.}
\end{deluxetable*}

\subsection{Photometry of the \arc}  \label{sec:arc_phot}

Our source extraction algorithm initially segmented the \arc\ into
several components, which we then merged into a single source in the
segmentation map (Figure~\ref{fig:arc}). We measured the photometry of
the \arc\ in all the observed HST and JWST filters using the \photutils\
SourceCatalog class as described in Section~\ref{sec:photcat}. As
before, we measured the photometry of the \arc\ in both an elliptical
Kron aperture with a scale factor of 1.5 and applied an aperture
correction to the fluxes and uncertainties for all filters to compute
total Kron fluxes. The size and shape of the elliptical Kron aperture
are determined by the central moments of the flux distribution of the
\arc\ in the detection image. This elliptical Kron aperture is shown in
Figure~\ref{fig:arc}. We also calculate isophotal fluxes by summing the
fluxes within the combined source segments defined by the segmentation
image. No additional correction factors were applied, and no additional
local background was subtracted.

For comparison, we also point spread function (PSF)-match our multiband
HST and JWST images to account for variations in the PSF between
filters. We extract empirical PSFs for the HST and JWST filters by
stacking images of bright, isolated stars in the data using the
\photutils\ package \citep{photutils220}. Using these PSFs, we construct
convolution kernels to homogenize the PSF sizes across all filters to
match that of the F160W filter, which has the broadest PSF. To assess
the reliability of the kernels, we convolve the PSFs of the other
filters with the corresponding kernels and compare their encircled
energy profiles to that of F160W, following the method presented in
\cite{Abdurrouf2023}. The profiles show excellent agreement, with
deviations smaller than 0.1 dex, confirming the robustness of our
PSF matching procedure. We find no significant difference between
the photometry derived from the PSF-matched images and our original
non-PSF-matched measurements.

The measured isophotal and Kron photometry of the \arc\ in the observed
HST and JWST filters is summarized in Tables~\ref{tbl:hstphot} and
\ref{tbl:jwstphot}. For completeness, we include the photometry
measured for both the non-PSF-matched and PSF-matched images. The arc
is brightest in the JWST NIRCam F200W image, with an observed magnitude
of $\mAB = 24.5$ in the Kron aperture (including the correction
to total flux) and $\mAB = 24.7$ in the segments. Each of the two
lensed mirror images is $\mAB = 25.3$, similar to that observed
for the brightest lensed image of MACS0647--JD at \zspec=10.17
\citep{Hsiao2024nirspec} with an observed F200W magnitude of $\mAB =
25.1$ \citep{Hsiao2023}. With a more modest magnification $\mu \sim 8$,
MACS0647--JD is intrinsically brighter ($\mAB = 27.3$ delensed) than the
\arc\ (see Section~\ref{sec:arc_mags}). MACS0647--JD is also observed
to be slightly brighter when adding the flux from all three lensed
images: $\mAB = 24.2$, compared to the \arc\ with $\mAB = 24.5$ (the
counterimage contributing negligibly).

For comparison, the exceptionally luminous unlensed galaxy GN-z11
at \zspec=10.60 \citep{Bunker2023} has an observed magnitude of
$\mAB = 26.0$ in the NIRCam F200W data \citep{Tacchella2023}. With
a few exceptions of lensed galaxies, most other galaxy candidates
at $z > 10$ being discovered by JWST are several magnitudes fainter
\citep[e.g.,][]{Finkelstein2024}.

\begin{figure*}[t!]
\begin{center}
\includegraphics[width=0.49\textwidth]{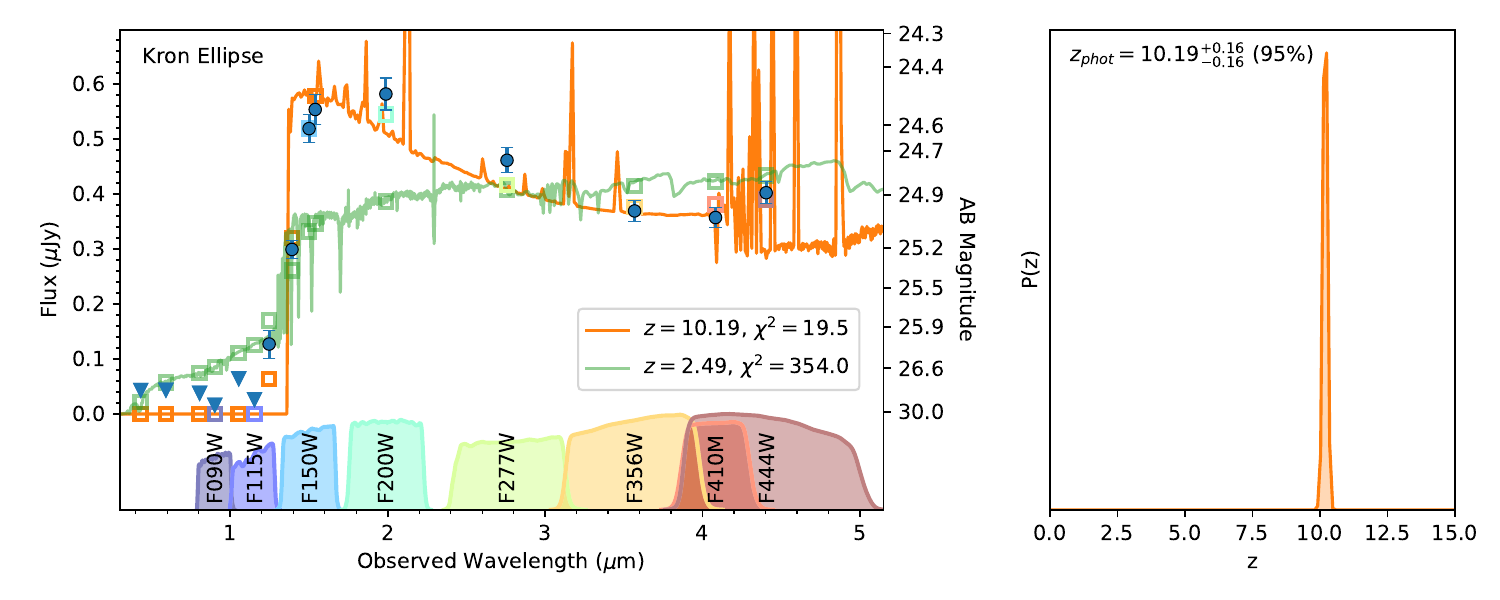}
\includegraphics[width=0.49\textwidth]{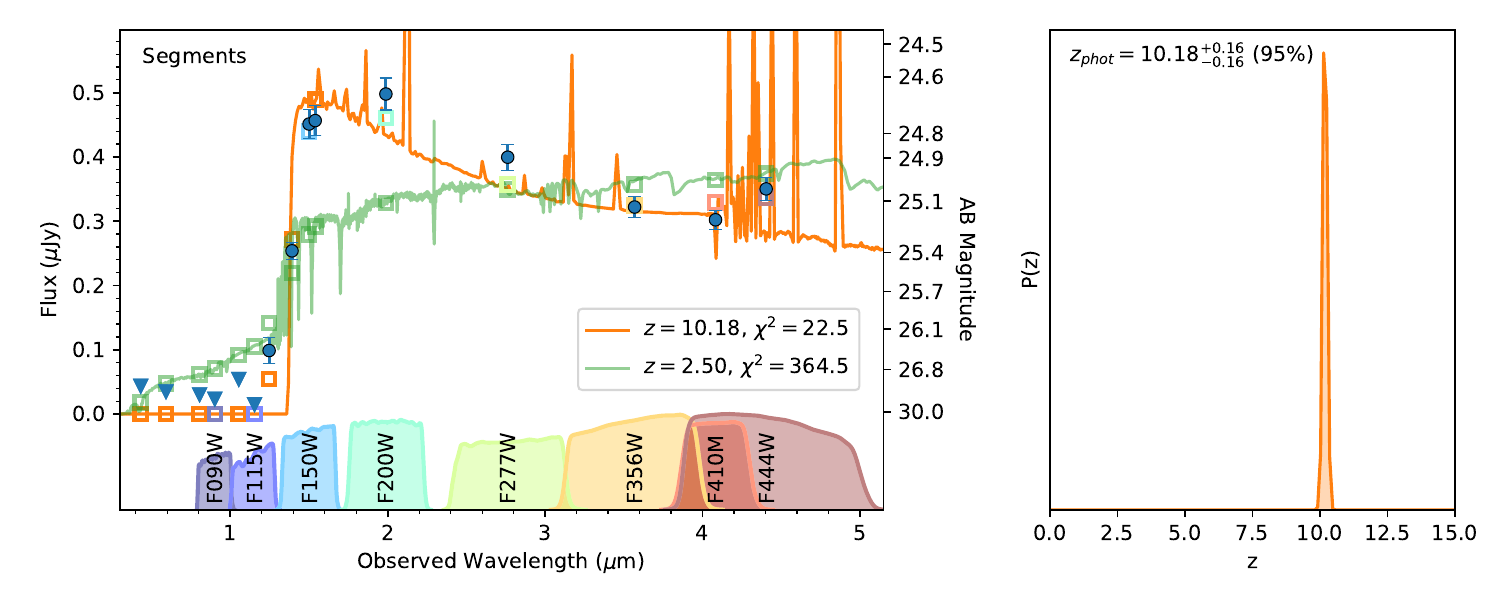}
\includegraphics[width=0.49\textwidth]{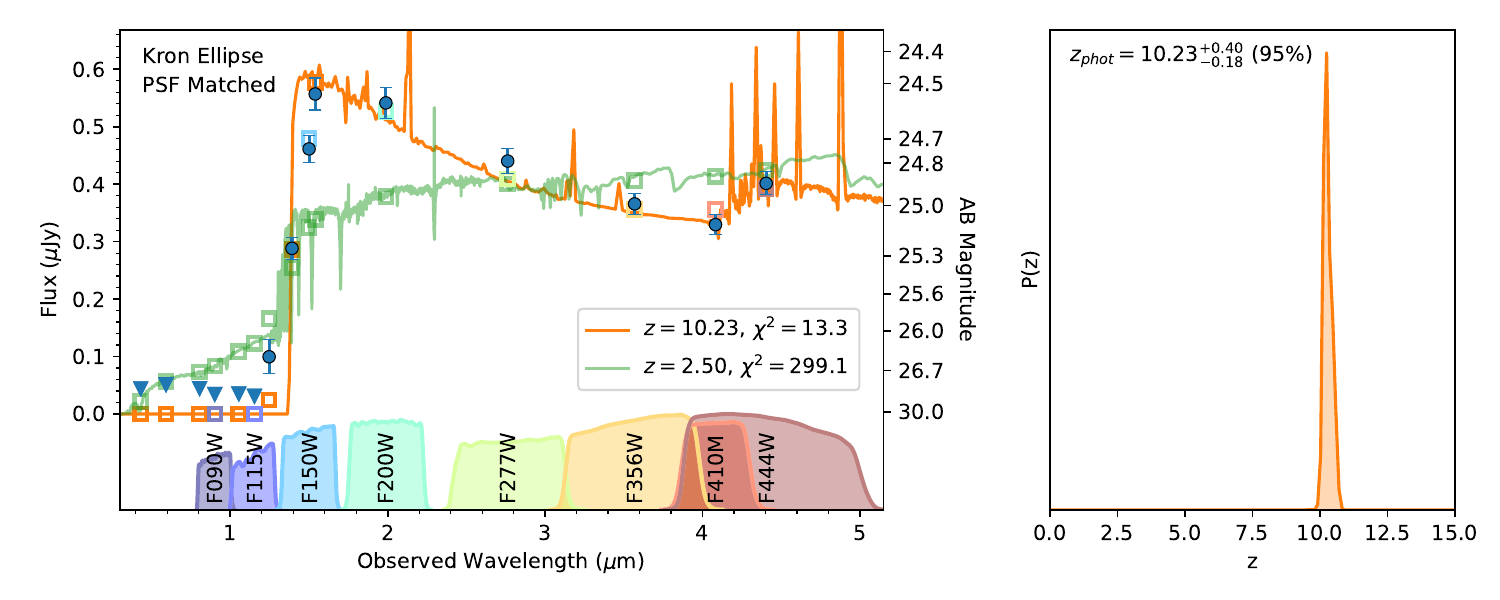}
\includegraphics[width=0.49\textwidth]{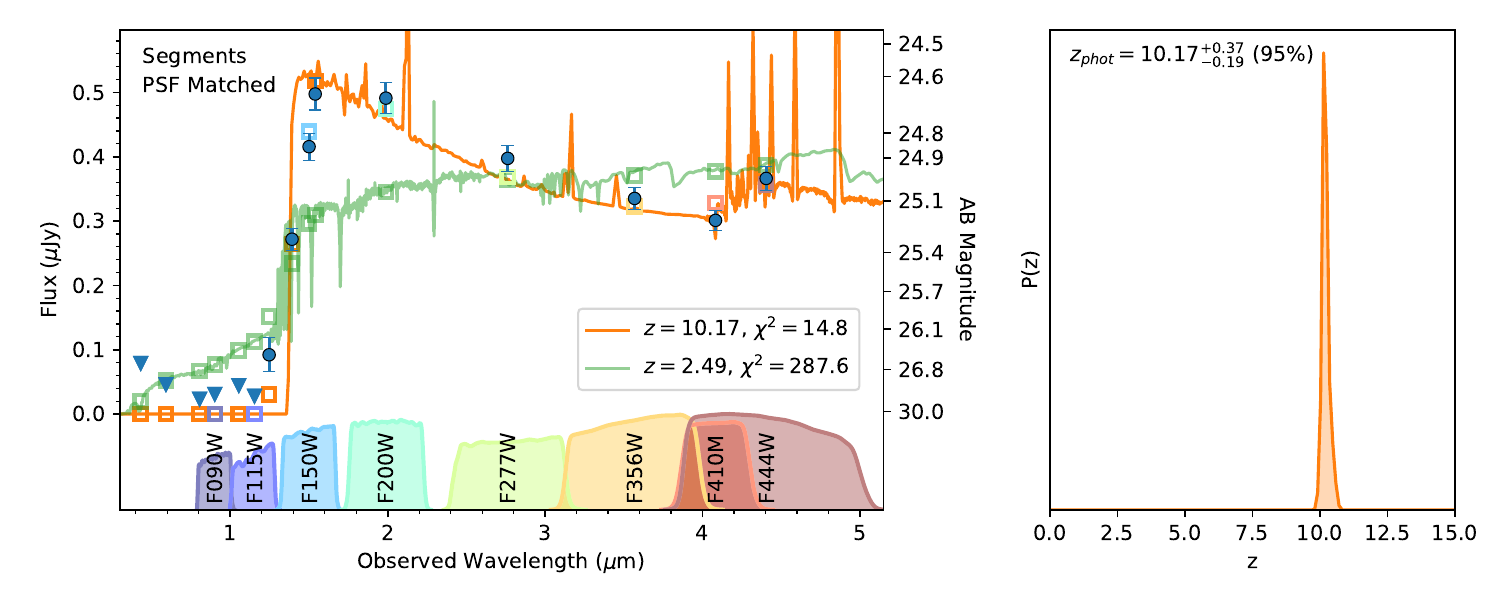}
\end{center}
\caption{
Best-fit \eazypy\ photometric redshifts for the \arc\ using the
photometry measured in both elliptical Kron apertures and isophotal
source segments. The top plots show results using the non-PSF-matched
photometry, while the bottom plots are for the PSF-matched photometry.
The measured fluxes are shown as blue data points or triangle upper
limits. Nondetections are plotted as upper limits at the $2\sigma$
level. The best-fit \eazypy\ SED model is shown in orange, with squares
indicating the expected photometry in a given band. The best-fit
\eazypy\ SED for the low-redshift ($z < 7$) solution is shown in green.
To the right of each SED, we also plot the P($z$) posterior redshift
probability distributions. The \arc\ has a best-fit photometric redshift
of $\zphot = 10.2$, with low-redshift solutions ruled out at high
significance.
\label{fig:arc_photoz}
}
\end{figure*}

\subsection{Photometric Redshifts}  \label{sec:arc_photoz}

The \arc\ has an extremely strong F115W$-$F200W break of $>3.2$~mag
($2\sigma$ lower limit) and is completely undetected ($< 2\sigma$) in
all bluer filters. These properties, combined with a very blue continuum
redward of the break, are consistent with a high-redshift Lyman-break
galaxy at $z \gtrsim 10$.

We measure the photometric redshift for the \arc\ using \eazypy\ (see
Section~\ref{sec:photoz}) from two separate sets of photometry: one
extracted from the elliptical Kron aperture and the other from the
isophotal segments. The results yield a photometric redshift of $\zphot
= 10.19 \pm 0.16$ (95\%) in the elliptical Kron aperture and $\zphot
= 10.18 \pm 0.16$ (95\%) in the segments. As a check, we remeasured
the photometric redshift using the PSF-matched photometry. This yields
$\zphot\ = 10.23_{-0.18}^{+0.40}$ (95\%) (using an elliptical Kron
aperture) and $\zphot\ = 10.17_{-0.19}^{+0.37}$ (95\%) (using the
segmentation map). While both methods produce a consistent best-fit
redshift of $\zphot \sim 10.2$, the PSF-matched photometry results in
a larger upper uncertainty (\zphot\ = $10.2_{-0.2}^{+0.4}$) compared
to the non-PSF-matched result (\zphot\ = $10.2_{-0.2}^{+0.2}$). Since
the PSF-matched results are consistent with our primary measurement, we
adopt the non-PSF-matched photometry for the remainder of our analysis.

We present the best-fit SED templates and the posterior redshift
distributions, P($z$), in Figure~\ref{fig:arc_photoz}. We also show
the best-fitting SED template where we restrict the redshift to be
$0.972 < z < 7$. We restrict the redshift to $z > 0.972$ as the arc is
clearly lensed and thus must lie behind the galaxy cluster. The best-fit
low-redshift models attempt to fit the very strong observed spectral
break with a Balmer break at $z \sim 2.5$, but are unable to match the
observed photometry. In all cases, the best-fit $\chi^2$ for these
``low-redshift'' solutions is significantly worse than the best-fit
high-redshift solutions, ruling out the low-redshift models at high
significance. Formally, 97\% of the P($z$) probability distribution
function is at $z > 10$ and increases to 100\% for $z > 9.8$. The
photometric redshift measurements of the \arc\ are summarized in
Table~\ref{tbl:hstphot}.

While the photometric redshift of the \arc\ is robust, definitive
confirmation of the redshift requires spectroscopic observations.
In 2023 December, we also obtained JWST NIRSpec high-resolution
G395H/F290LP spectroscopy (8928~s total) from the same JWST Cycle 2
program (JWST-GO-4212) in several multi-shutter array slitlets along
the \arc. The primary goals of these observations were to obtain
a spectroscopic redshift of the galaxy and to detect and resolve
the \OIIww\ doublet, measuring electron densities along the arc.
Incidentally, the first such detection of the resolved \OIIww\ doublet
at $z > 8$ was recently made from NIRSpec G395H observations of the
lensed galaxy MACS0647-JD at $z=10.17$ \citep{Abdurrouf2024}.

Unfortunately, the G395H spectroscopic data does not reveal any
emission lines over the wavelength range of the observations (2.9 --
5.3~\micron), which covers the rest-frame of 2636 -- 4818~\AA\ at $z
= 10.2$. The continuum was also not detected, as expected for the
high-resolution G395H grating setting. Our team was recently awarded
JWST cycle 3 (PI: Vanzella) observing time that includes NIRSpec IFU
prism observations of the \arc, which will be used to confirm the
redshift and study the arc in more detail \citep{Messa2025}.

\subsection{Magnifications}  \label{sec:arc_mags}

All four of our independent lens models produce excellent results
consistent with the $z=10.2$ critical curve crossing the \arc,
which confirms that the observed symmetry of the arc is indeed the
result of seeing two lensed mirror images of the galaxy (denoted in
Figure~\ref{fig:arc}). The $z=10.2$ critical curve for our fiducial
\lta\ model is shown in Figure~\ref{fig:cluster} overlaid on the \SPT\
cluster image. The \arc\ inset image (Figure~\ref{fig:cluster}) shows
the critical curves from all four lens models bisecting the arc.

To verify that our results are not biased by a misinterpretation of the
lensing symmetry of the arc, we ran a consistency check by excluding
the positional constraints of the \arc\ from the lens models. For the
\glafic\ and \lta\ lens models, even without this constraint, the
critical curve still passes through the center of the \arc, confirming
our interpretation of it being a pair of multiple images.

For the \wslap\ lens model, when the $z=10.2$ arc is not included as
a constraint, the model predicts the critical curve passing $\sim
1\arcsec$ from the \arc. When the arc is included as a constraint, the
predicted critical curve passes between the star clusters C.1 and D.1,
just 0\farcs3 from the alleged symmetry point in the arc and within the
uncertainties typical of \wslap\ models.

While the star clusters in the arc appear overall to be symmetric,
the distance between A.1 and B.1 is greater than the distance between
A.2 and B.2 by a factor of $\sim1.4$. The appearance of Image 2 is
likely affected by additional lensing effects from the \zphot = 2.6 arc
immediately to the north of the \arc\ (see Figure~\ref{fig:cutouts}). We
investigated this possibility by running a \wslap\ model that includes
the $z=2.6$ arc. While the agreement improves, it is not yet sufficient
to completely explain the observed perturbation in the arc. Adding
at least one other small-scale dwarf galaxy near the \arc\ below our
current detection limit could explain the perturbed magnification of
Image 2. We are exploring this possibility with a new \wslap\ model that
includes additional perturbers, but defer such analysis to future work.

Because the arc lies on the critical curve, it also has a very large
magnification. The arc is highly magnified in the tangential direction
along the arc, with only modest magnification in the radial direction
($\mu \sim 1.3$). The magnifications also vary along the arc, with
higher values near the critical curve, and are consistent among our lens
models (see values reported in \cite{Adamo2024} at the star cluster
positions). For our reference \lta\ model, the star cluster E.1 has the
highest magnification of $\mu = 419$, while the star cluster A.1 has the
lowest magnification of $\mu = 57$.

The median magnification of each mirrored image is $\mu \sim
60_{-8}^{+17}$ (95\% credible interval). To estimate the uncertainties,
we generated 100 magnification maps by sampling the MCMC posterior
distribution of our fiducial \lta\ model. For each map, we measure the
median magnification along the arc and estimate the 95\% confidence
interval on each side of the best magnification value of $\mu = 60$ to
obtain asymmetrical errors. We note that our calculated magnification
errors are likely underestimated, as they do not include the impact of
systematic uncertainties near the critical curve. In this region of
extreme magnification sensitivity, a small change in the assumed mass
distribution of the lens can move the position of this curve, causing a
large magnification change. Assuming that we are observing two lensed
images of the same galaxy, in total, the portion of the galaxy that is
lensed to form the \arc\ is magnified by $\mu \sim 120_{-16}^{+34}$. We
use this value to compute the intrinsic properties of the whole galaxy
from the measured values.

The \arc\ has an intrinsic F200W apparent magnitude of $\mAB =
29.7$. This corresponds to an absolute magnitude of $\MAB = -17.8$
in the rest-frame UV. This makes the \arc\ less luminous than the
typical $M^{*}$ galaxy at $z \sim 10$ \citep[e.g.,][]{Adams2024,
Finkelstein2024}, placing it in the category of galaxies likely
to have driven cosmic reionization. This is especially true if
the escape fraction of ionizing radiation from compact star
clusters is close to unity, as suggested by some theoretical models
\citep[see][]{Ricotti2002, HeR2020}.

\begin{figure*}[t!]
\begin{center}
\includegraphics[width=0.8\textwidth]{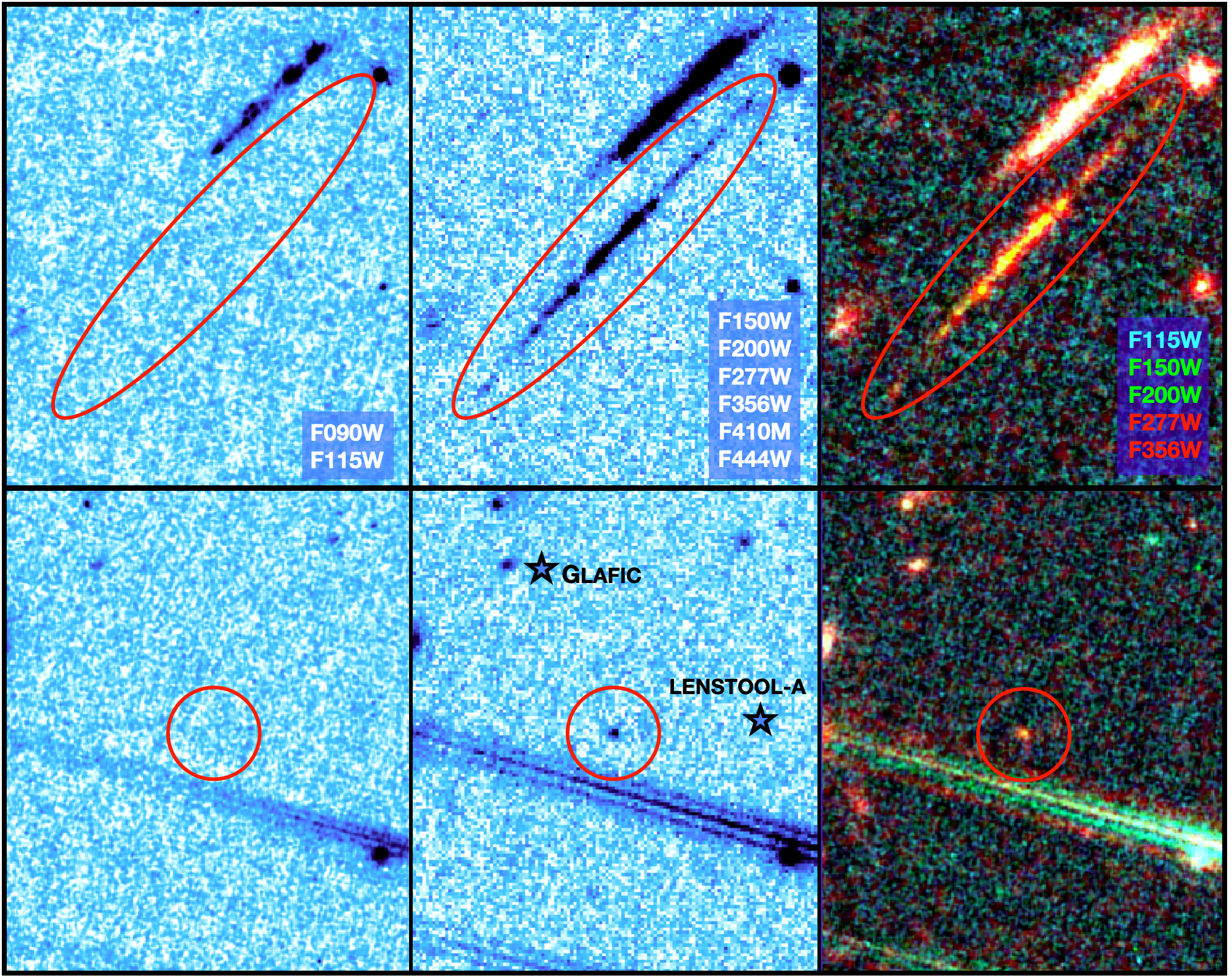}
\end{center}
\caption{
Cutout images of the \arc\ (top) and the candidate counterimage
(bottom), showing the \Lya-break using the JWST bands only. The field
of view of the cutouts is $5\arcsec \times 6\arcsec$, and the images are
shown with north up and east left. The stacked images blueward (F090W
+ F115W) and redward (F150W + F200W + F356W + F410M + F444W) of the
\Lya-break are shown in the left and central panels, respectively. The
right panels show color composites in the NIRCam filters. The \arc\ has
an extremely strong NIRCam F115W$-$F200W break of $>3.2$~mag ($2\sigma$
lower limit), is undetected ($< 2\sigma$) in all bluer filters, and
has a very blue continuum slope redward of the break. The candidate
counterimage has similar colors to the arc, but is 3.9~mag fainter, with
an observed F200W magnitude of $\mAB = 28.4$, fully consistent with the
\lta\ model prediction of the counterimage being $\sim 3.7 - 4.7$~mag
fainter. The predicted locations of the counterimage from the \lta\ and
\glafic\ models (star symbols in the bottom-center panel) are within
1\farcs8 and 2\farcs2 of the candidate counterimage, respectively.
\label{fig:cimg_cutouts}
}
\end{figure*}

\begin{figure*}[t!]
\begin{center}
\includegraphics[width=0.98\textwidth]{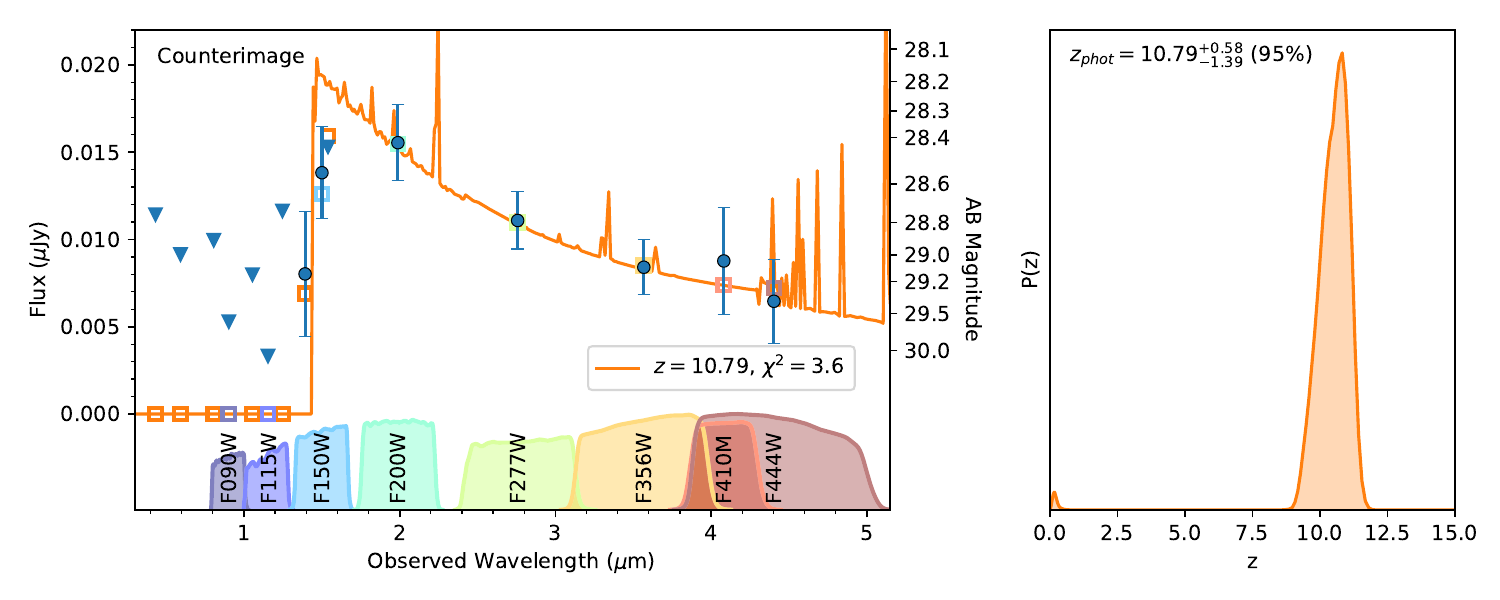}
\end{center}
\caption{
Best-fit \eazypy\ photometric redshift for the candidate counterimage.
The measured fluxes are shown as blue data points or triangle upper
limits. Nondetections are plotted as upper limits at the $2\sigma$
level. The best-fit \eazypy\ SED model is shown in orange, with squares
indicating the expected photometry in a given band. In the right-hand
panel, we plot the P($z$) posterior redshift probability distribution.
The candidate counterimage has a best-fit photometric redshift of
$\zphot = 10.8_{-1.4}^{+0.6}$, which is consistent with the redshift of
the \arc\ of $\zphot = 10.2 \pm 0.2$.
\label{fig:cimg_eazy}
}
\end{figure*}

\subsection{Candidate Counterimage}

All four of our independent lens models predict a fainter counterimage
of the \arc\ at similar locations. The reference \lta\ model predicts
the counterimage at (R.A., decl.) = ($93.9490607$, $-57.7701814$).
The \ltb\ predicted location is $\sim$2\arcsec\ away. The \glafic\
best model prediction is also nearby at (R.A., decl.) = ($93.9504865$,
$-57.7696559$). The \wslap\ prediction is consistent with the other
models. All the models predict a much lower magnification for the
counterimage than the \arc, in the range of $\mu\sim1.5 - 2.0$.

We identify a potential candidate counterimage with the same colors
of the \arc\ at (R.A., decl.) = ($93.95000245$, $-57.77021846$), as
shown in Figure~\ref{fig:cluster}. This is within $\sim 1\farcs8$ and
2\farcs2 of the predictions from \lta\ and \glafic, respectively,
as shown in Figure~\ref{fig:cimg_cutouts}. The measured photometry
of the candidate counterimage in the observed HST and JWST filters
is summarized in Tables~\ref{tbl:hstphot} and \ref{tbl:jwstphot}.
The candidate counterimage is 3.9~mag fainter than the \arc, with
an observed F200W magnitude of $\mAB = 28.4$, consistent with the
expectation from the lens models. Earlier lens models based only on HST
data \citep{Paterno-Mahler2018, Salmon2018} predicted a counterimage in
proximity to our candidate, but those works were unable to identify the
counterimage at the shallower depth of the HST data.

Assuming the \lta\ magnification estimate of $\mu = 2.0 \pm 0.1$ for the
counterimage yields a delensed magnitude $\mAB = 29.2$, which is 60\%
brighter than the delensed magnitude estimate for the \arc\ of $\mAB
= 29.7$. This may be explained by the portion of the galaxy missing
in the fold of the lensing critical curve in the \arc\ image, whereas
the counterimage is a complete image of the source galaxy. Lens model
magnification uncertainties may contribute as well.

We measure a photometric redshift of $z_{phot} = 10.8_{-1.4}^{+0.6}$
for the candidate counterimage, which is consistent with the redshift
of the \arc\ given the larger uncertainties. We present its best-fit
\eazypy\ SED model and the posterior redshift distribution in
Figure~\ref{fig:cimg_eazy}.

Finally, we note that according to the \lta\ model, each star cluster
is magnified $\sim 3.7 - 4.7$~mag brighter in the \arc\ compared
to the counterimage. Thus, with intrinsic magnitudes $\sim 31$ and
fainter, none of the star clusters can be individually detected in the
counterimage. Nor would they be detected without lensing in a blank
field. They are only discernible thanks to the very strong lensing of
the \arc.

\tabletypesize{\small}
\begin{deluxetable*}{cccccccc}
\tablecolumns{8}
\tablecaption{\bagpipes\ photometric redshifts and physical properties of \sptarc
\label{tbl:bagpipes_properties}}
\tablehead{
\colhead{Source} &
\colhead{\zphot\tnm{a}} &
\colhead{\logmstar} &
\colhead{SFR\tnm{b}} &
\colhead{$\log$ sSFR/Gyr$^{-1}$} &
\colhead{Age\tnm{c}} &
\colhead{$A_V$} &
\colhead{\tform\tnm{d}}
\\[-0.5em]
\colhead{} &
\colhead{} &
\colhead{} &
\colhead{($\Msun$ \peryr)} &
\colhead{} &
\colhead{(Myr)} &
\colhead{(mag)} &
\colhead{(Myr)}
}
\startdata
\sidehead{Delayed-$\tau$ SFH model, SMC dust extinction}
\hline
Kron Ellipse & $10.2_{-0.1}^{+0.1}$ & $7.47_{-0.18}^{+0.18}$ & $0.33_{-0.09}^{+0.03}$ & $1.05_{-0.14}^{+0.03}$ & $49_{-22}^{+40}$ & $0.02_{-0.01}^{+0.01}$ & $406_{-41}^{+24}$ \\
Segments & $10.2_{-0.1}^{+0.1}$ & $7.38_{-0.21}^{+0.18}$ & $0.28_{-0.10}^{+0.03}$ & $1.06_{-0.13}^{+0.03}$ & $44_{-22}^{+39}$ & $0.03_{-0.01}^{+0.01}$ & $411_{-40}^{+24}$ \\
\hline
\sidehead{Delayed-$\tau$ SFH model, SMC dust extinction, BPASS stellar models}
\hline
Kron Ellipse & $10.1_{-0.1}^{+0.1}$ & $7.53_{-0.19}^{+0.12}$ & $0.28_{-0.03}^{+0.01}$ & $0.93_{-0.12}^{+0.13}$ & $78_{-36}^{+36}$ & $0.01_{-0.01}^{+0.01}$ & $380_{-36}^{+35}$ \\
Segments & $10.1_{-0.1}^{+0.1}$ & $7.47_{-0.17}^{+0.12}$ & $0.24_{-0.02}^{+0.01}$ & $0.93_{-0.12}^{+0.13}$ & $79_{-34}^{+38}$ & $0.01_{-0.01}^{+0.01}$ & $378_{-37}^{+34}$ \\
\hline
\sidehead{Delayed-$\tau$ SFH model, Calzetti dust extinction}
\hline
Kron Ellipse & $10.2_{-0.1}^{+0.1}$ & $7.49_{-0.20}^{+0.17}$ & $0.34_{-0.11}^{+0.04}$ & $1.05_{-0.14}^{+0.03}$ & $47_{-23}^{+41}$ & $0.06_{-0.03}^{+0.03}$ & $407_{-40}^{+25}$ \\
Segments & $10.2_{-0.1}^{+0.1}$ & $7.40_{-0.22}^{+0.20}$ & $0.29_{-0.11}^{+0.05}$ & $1.06_{-0.13}^{+0.02}$ & $40_{-21}^{+43}$ & $0.07_{-0.03}^{+0.03}$ & $415_{-45}^{+22}$ \\
\hline
\sidehead{Exponential SFH model, SMC dust extinction}
\hline
Kron Ellipse & $10.2_{-0.1}^{+0.1}$ & $7.45_{-0.13}^{+0.17}$ & $0.35_{-0.10}^{+0.07}$ & $1.08_{-0.08}^{+0.02}$ & $36_{-13}^{+30}$ & $0.03_{-0.01}^{+0.01}$ & $419_{-30}^{+14}$ \\
Segments & $10.2_{-0.1}^{+0.1}$ & $7.39_{-0.15}^{+0.18}$ & $0.30_{-0.10}^{+0.06}$ & $1.08_{-0.10}^{+0.02}$ & $36_{-14}^{+32}$ & $0.03_{-0.01}^{+0.01}$ & $419_{-34}^{+16}$ \\
\hline
\sidehead{Constant SFH model, SMC dust extinction}
\hline
Kron Ellipse & $10.1_{-0.1}^{+0.1}$ & $7.75_{-0.09}^{+0.05}$ & $0.68_{-0.13}^{+0.09}$ & $1.08_{-0.01}^{+0.01}$ & $22_{-4}^{+5}$ & $0.14_{-0.04}^{+0.03}$ & $435_{-8}^{+7}$ \\
Segments & $10.1_{-0.1}^{+0.1}$ & $7.68_{-0.10}^{+0.06}$ & $0.58_{-0.12}^{+0.09}$ & $1.08_{-0.01}^{+0.01}$ & $21_{-4}^{+6}$ & $0.15_{-0.04}^{+0.03}$ & $436_{-8}^{+7}$
\enddata
\tablecomments{Properties are quoted as the median and the 68\% range of the joint posterior distributions. Stellar masses and SFRs are corrected for magnification assuming a total magnification of $\mu \sim 120_{-16}^{+34}$. Multiply these values by 120 / $\mu$ to apply a different magnification. We did not propagate magnification uncertainties to those parameter uncertainties. Applying a range of magnifications within the uncertainties would increase or decrease the stellar masses and SFRs by an additional 15\% or 22\%, respectively.
}
\tablenotetext{a}{Photometric redshift with 95\% confidence interval.}
\tablenotetext{b}{SFR during the past 100 Myr.}
\tablenotetext{c}{Mass-weighted age.}
\tablenotetext{d}{Formation time in Myr after the Big Bang based on the mass-weighted age.}
\end{deluxetable*}
\tabletypesize{\normalsize}

\begin{figure*}[t!]
\begin{center}
\includegraphics[width=0.49\textwidth]{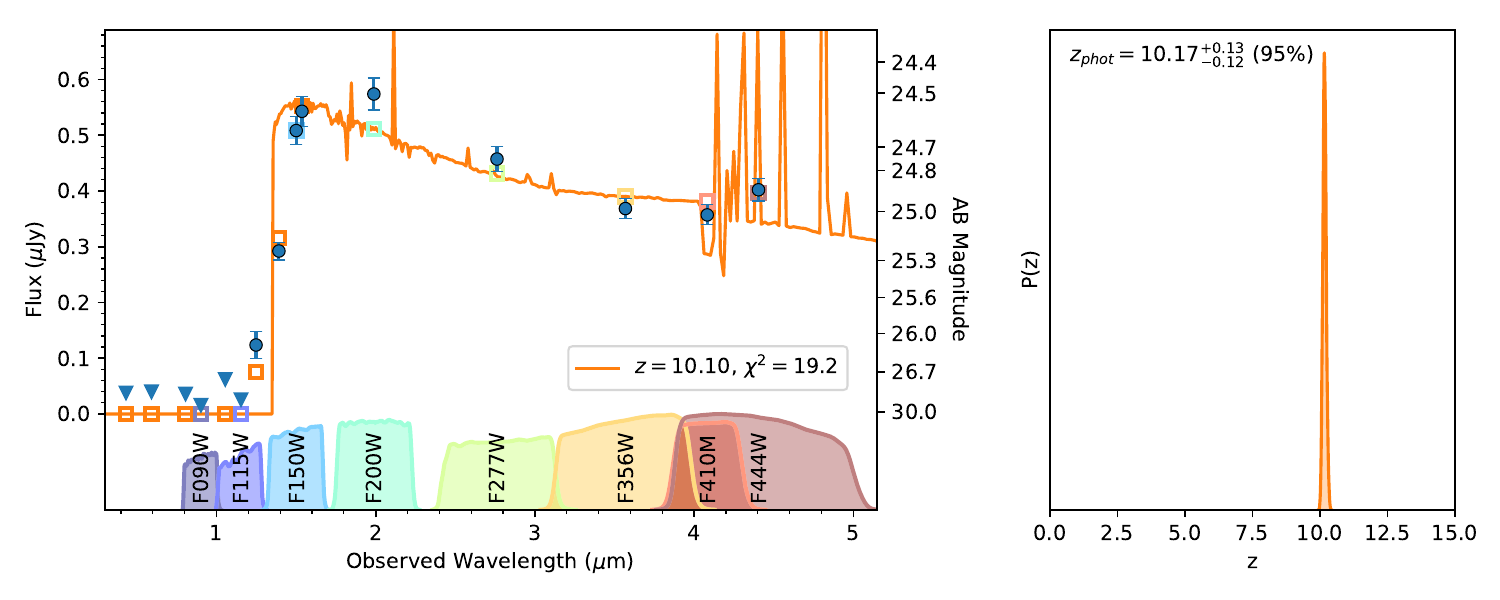}
\includegraphics[width=0.49\textwidth]{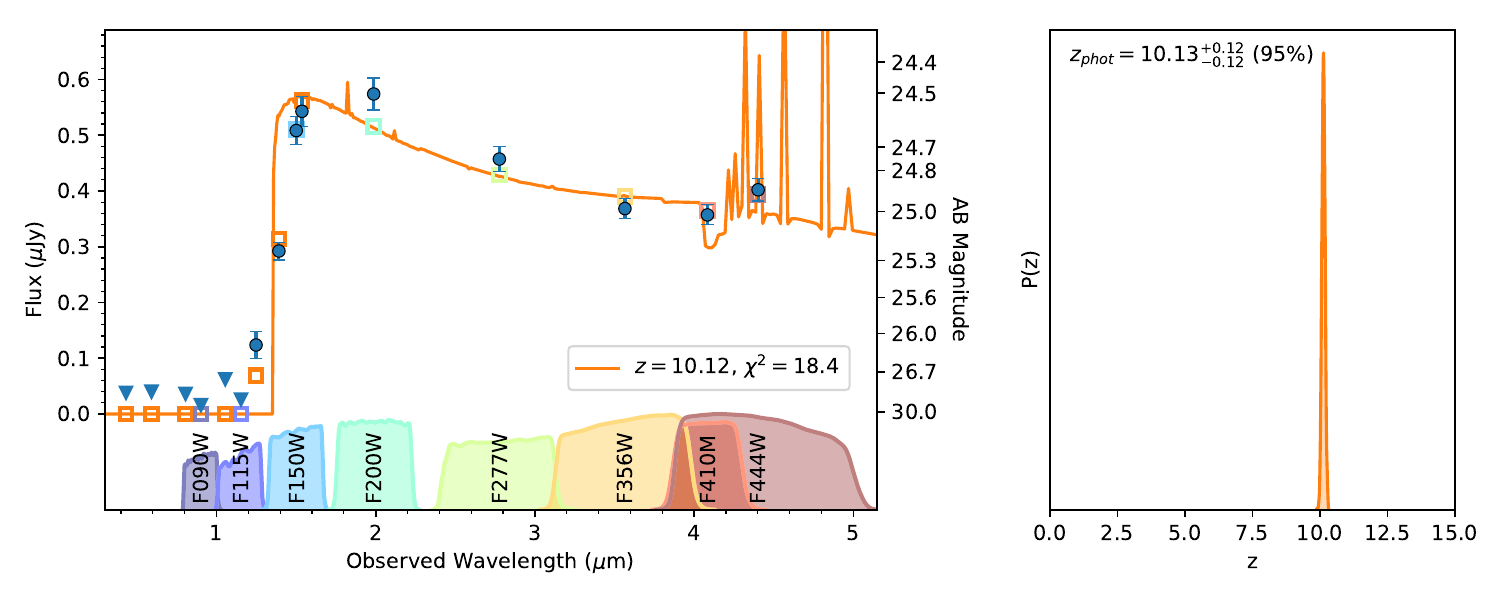}
\includegraphics[width=0.49\textwidth]{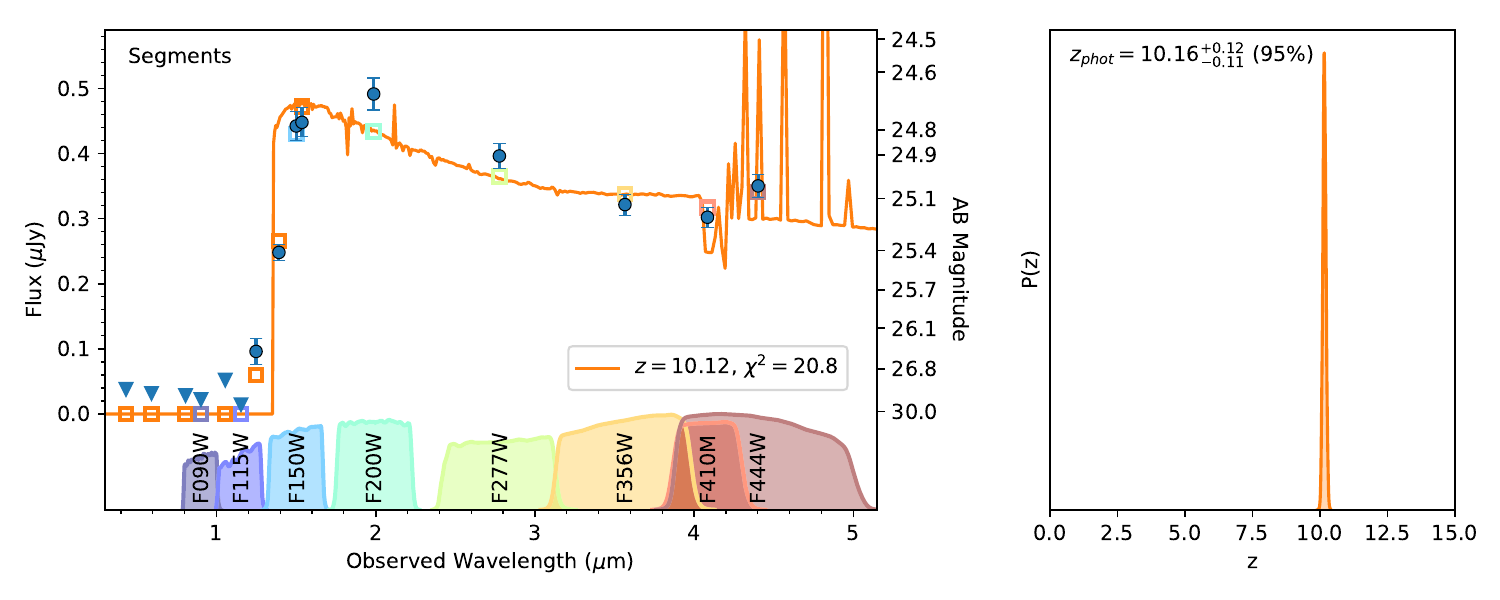}
\includegraphics[width=0.49\textwidth]{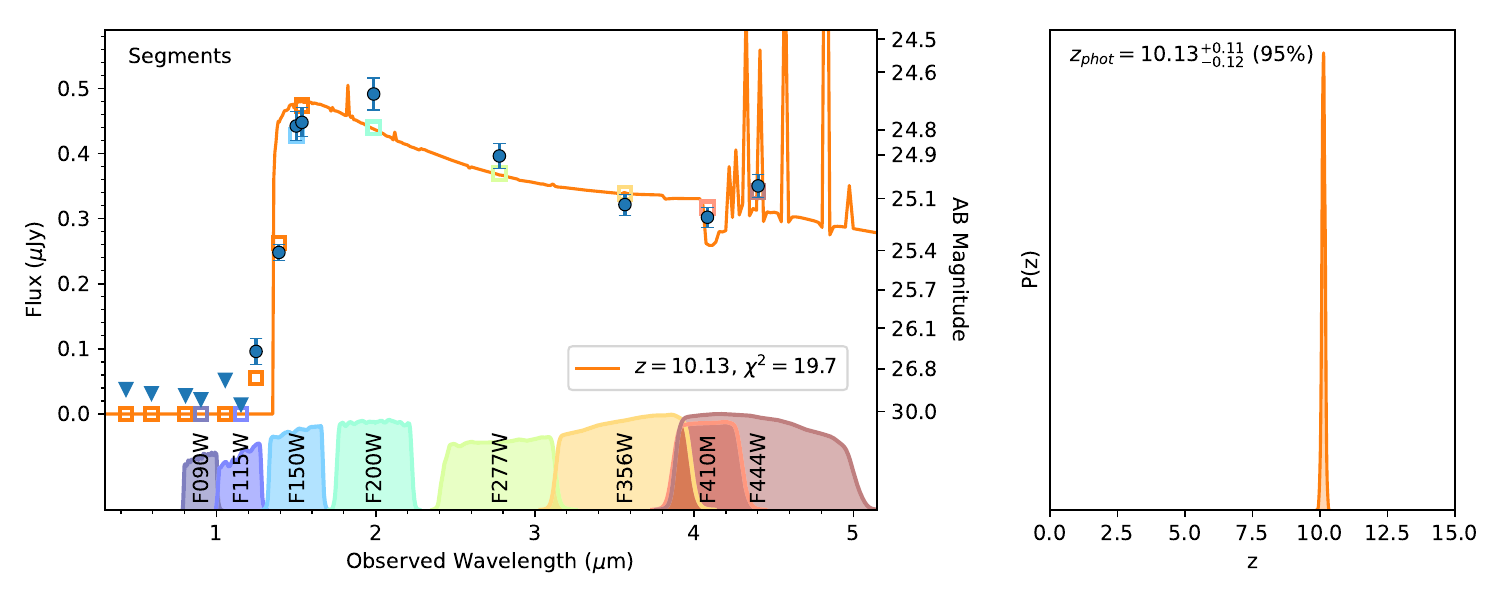}
\end{center}
\caption{
Best-fit \bagpipes\ SED models for the \arc\ assuming a delayed-$\tau$
SFH and a SMC dust extinction law. The top panels show the fits to
fluxes measured in the Kron Ellipse, while the bottom panels show
the fits to segment photometry. The left panels show the best fits
using the default BC03 stellar population models, while the right
panels show the best fits using the BPASS stellar population models.
The measured fluxes are shown as blue data points or triangle upper
limits. Nondetections are plotted as upper limits at the $2\sigma$
level. The best-fit \bagpipes\ SED model is shown in orange, with
squares indicating the expected photometry in a given band. To the right
of each SED, we also plot the P($z$) posterior redshift probability
distributions with the median \zphot\ and 95\% confidence interval.
\label{fig:bagpipes}
}
\end{figure*}

\subsection{Physical Properties} \label{sec:properties}

The physical properties of the \arc\ were estimated using the Bayesian
Analysis of Galaxies for Physical Inference and Parameter EStimation
(\bagpipes) SED-fitting code \citep{bagpipes}. \bagpipes\ fits
the observed photometry of a galaxy by generating model galaxy
spectra over the multidimensional space of physical parameters. The
fits are performed using the \multinest\ nested sampling algorithm
\citep{Feroz2009}.

By default, \bagpipes\ uses the stellar population synthesis models
from the 2016 version of the BC03 \citep{BC03} models with a
\cite{Kroupa2002} IMF. These models also include nebular line and
continuum emission based on \cloudy\ \citep{Ferland2013} with the
logarithm of ionization parameter ($\log$~U) allowed to vary between
$-4$ and $-2$. We vary metallicity in logarithmic space from $\log
Z/\Zsun = -4$ to $-0.7$. Formation ages vary from 1 Myr to the age of
the Universe.

We also explore the effect of using the BPASS stellar population models
\citep{Eldridge2009} on the derived physical properties. The BPASS
models include binary stellar evolution and account for the effects of
binary interactions on the stellar population \citep{Eldridge2017}. We
use the BPASS v2.2.1 models \citep{Stanway2018} with an IMF slope of
$-1.30$ for stars with masses $0.1 - 0.5~\Msun$, $-2.35$ for stars with
masses $0.5 - 300~\Msun$, and a maximum stellar mass of $300~\Msun$.

We explore several SFHs to estimate the physical properties of the \arc.
The SFHs include delayed exponentially declining $\tau$ models, constant
star formation rate (SFR) models, and exponentially declining models.
For the delayed-$\tau$ model, the SFR is of the form SFR$(t) \propto t
\ \exp{(-t/\tau)}$, where $\tau$ is the e-folding time. This SFR rises
linearly with time and then exponentially declines if $\tau$ is less
than the formation age.

We generally assume a Small Magellanic Cloud (SMC) dust extinction law
\citep{Salim2018}. However, we also investigate the effect of using
the Calzetti extinction law \citep{Calzetti2000}. For both extinction
laws, we also add a second component to the dust model that includes
birth-cloud dust attenuation. This attenuation is a factor of 2 larger
around \HII\ regions, as in the general ISM within the galaxy's first 10
Myr. We allow dust extinction to range from $A_V$ = $0 - 1$ mag.

We fit the observed photometry using the Kron ellipse and the isophotal
segments of the \arc. As with the \eazypy\ fits, we apply an error floor
of 5\% to the flux uncertainties to account for photometric calibration
uncertainties. The redshifts are allowed to span from $9.9 < z < 10.5$
(i.e., $\pm 3\sigma$) based on the measured \eazypy\ $\zphot = 10.2 \pm
0.2$ (95\%).

The physical properties of the \arc\ are presented in
Table~\ref{tbl:bagpipes_properties}. The reported stellar masses and
SFRs are corrected for magnification and assume that we are observing
two mirror images of a galaxy, as suggested by the lens models (see
Section~\ref{sec:arc_mags}). In Figure~\ref{fig:bagpipes}, we show the
best-fit \bagpipes\ SED models for the \arc, fitting both the fluxes
from the Kron Ellipse and segments. Fits for both are shown for the
default BC03 models and BPASS models.

For most \bagpipes\ models, the best-fit photometric redshift for
the \arc\ is $\zphot = 10.2 \pm 0.1$, in agreement with the \eazypy\
photometric redshift of $\zphot = 10.2 \pm 0.2$. Some models yield a
slightly lower $\zphot = 10.1 \pm 0.1$, but all are consistent with the
\eazypy\ redshift.

The intrinsic stellar masses of the \arc\ are consistent among the
different models, with typical \logmstar\ values of $7.4 - 7.5$ ($\Mstar
= 2.5 - 3.2 \times 10^{7}~\Msun$). The constant SFR models yield
slightly larger intrinsic stellar masses of $\logmstar = 7.7 - 7.8$
($\Mstar = 4.7 - 5.6 \times 10^{7}~\Msun$). These stellar masses are
lower than the typical stellar masses of $10^{8} - 10^{9}~\Msun$
found for galaxies at $z \sim 10$ \citep{Finkelstein2022, Naidu2022a,
Bradley2023, Bunker2023, Hsiao2023, Stiavelli2023}. Applying a range of
magnifications within the uncertainties would increase or decrease the
stellar masses by an additional 15\% or 22\%, respectively.

The mass-weighted ages of the \arc\ indicate a relatively young stellar
population, with ages typically $\sim 36 - 49$~Myr. These galaxy ages
are slightly older than the ages derived for the individual star
clusters, which range from 9 to 35~Myr \citep{Adamo2024}, suggesting
they formed only a few tens of Myr after the galaxy and, indeed,
constitute a significant fraction of the galaxy mass \citep[$\sim
30$\%][]{Adamo2024}. The constant SFR models yield even younger ages of
$\sim 21 - 22$~Myr, while the BPASS models prefer older ages of $\sim 78
- 79$~Myr. The formation time of the \arc\ is $\tform = 378 - 436$~Myr
after the Big Bang.

Most models yield similar SFRs of $\sim 0.2 - 0.3~\Msun$~\peryr\
(averaged over the last 100~Myr), with the constant SFR models yielding
larger SFRs of $\sim 0.6 - 0.7~\Msun$~\peryr\ with corresponding
younger ages of only $\sim 22$~Myr. We note that the SFRs can larger or
smaller by an additional 15\% or 22\%, respectively, given the range
magnifications within the uncertainties. This same effect of younger
ages and larger SFRs for constant SFR models was observed for a sample
of four high-redshift galaxies at $z \sim 9 - 10$ by \cite{Bradley2023}.

We also find very low dust content with $A_V < 0.07$~mag for most
models. For the constant SFR models, the dust extinction is slightly
larger with $A_V = 0.14 - 0.15$~mag. The low dust content is consistent
with the very blue rest-frame UV slope ($\beta = -2.7 \pm 0.1$) measured
from a power-law fit to the F200W, F277W, and F356W photometry. These
results are also consistent with the low dust extinction found in the
individual star clusters from SED fitting \citep{Adamo2024}.

Finally, the best-fitting \bagpipes\ results suggest a very low
metallicity of $\lesssim 1\%~\Zsun$. While it is difficult to constrain
the metallicity with the current photometric data alone, the low
metallicity is consistent with the low dust content and young stellar
population inferred from the SED fitting. The low metallicity is also
consistent with the low metallicity ($\sim$ 5\%~\Zsun) found in the
individual star clusters of the \arc\ \citep{Adamo2024}. The inferred
low metallicity is interesting in the context of the reionization era
galaxies, as it suggests that the \arc\ may be a low-metallicity
star-forming galaxy.

The low dust content and low metallicity are also consistent with
the non-detection of the dust continuum and the \OIIIfir\ line
from deep Atacama Large Millimeter/submillimeter Array cycle
6 and 7 observations (PI: Tamura; program IDs 2018.1.00295.S
and 2019.1.00327.S) of the \arc\ covering $z = 9.5 - 10.7$. The
lack of any bright emission lines (including \OIIww) in the JWST
NIRSpec G395H MOS observations could also be explained by low
metallicity. The apparent excess of flux in the NIRCam F200W band
(see Figure~\ref{fig:bagpipes}) could be due to the presence of the
\HeIIw\ line, which is expected to be strong in low-metallicity or
Population III galaxies \citep[e.g.,][]{Nakajima2022}. If the \arc\ is
a low-metallicity or Population III galaxy, the \HeIIw\ line will be
clearly detected (EW $>$ 30~\AA) in the upcoming JWST Cycle 3 NIRSpec
IFU prism observations of the \arc.

\section{Conclusions} \label{sec:conclusions}

We present a detailed analysis of the \arc\ (\sptarc), a galaxy
candidate at $z \sim 10.2$, using a combination of recent JWST NIRCam
imaging data in eight bands and archival HST imaging data in seven
bands, in total spanning $0.4 - 5.0$~\micron. The \arc\ is the most
highly magnified ($\mu = 120_{-16}^{+34}$) and second-brightest observed
galaxy known at $z\gtrsim10$, magnified to 24.5~AB mag in the NIRCam
F200W band. The combination of magnified brightness and resolution makes
the \arc\ a unique laboratory to perform spatially resolved studies not
possible in any other galaxy at this distance.

The \arc\ extends 5\arcsec\ in the JWST NIRCam images. There is a
distinctive symmetry of bright knots from the center of the extended
arc, suggesting that the observed arc represents two mirror images
of the lensed galaxy. This is confirmed by our four independent
cluster lens models (see Section~\ref{sec:lens_models}), all of
which predict the $z=10.2$ critical curve crossing the arc (see
Figure~\ref{fig:cluster} inset). The bright knots in the \arc\ have been
identified as five individual star clusters, resolved down to 1~pc in
intrinsic size in the JWST NIRCam F150W image \citep{Adamo2024}. This
is the first such detection of star clusters at $z > 10$, only 460~Myr
after the Big Bang.

The arc has an extremely strong NIRCam F115W$-$F200W break of $>3.2$~mag
($2\sigma$ lower limit), is undetected ($< 2\sigma$) in all bluer
filters, and has a very blue continuum slope redward of the break. Using
\eazypy, the best-fit photometric redshift for the \arc\ is $\zphot
= 10.2 \pm 0.2$. This is in agreement with the best-fit \bagpipes\
photometric redshift of $\zphot = 10.2 \pm 0.1$. We find that the arc is
a low-luminosity galaxy ($\MUV = -17.8$) with a very blue rest-frame UV
slope ($\beta = -2.7 \pm 0.1$). Thus, the \arc\ is less luminous than
the typical $M^{*}$ galaxy at $z \sim 10$ \citep[e.g.,][]{Adams2024,
Finkelstein2024}, placing it in the category of galaxies likely to have
driven cosmic reionization.

We use the \bagpipes\ code to estimate the physical properties of the
\arc. Assuming a total magnification of $\mu = 120_{-16}^{+34}$, the
intrinsic stellar mass of the \arc\ is $\logmstar = 7.4 - 7.8$ ($\Mstar
= 2.4 - 5.6 \times 10^{7}~\Msun$), with a mass-weighted age of $\sim 21
- 79$~Myr. The SFR of the \arc\ is $\sim 0.2 - 0.7~\Msun$ \peryr, with
a very low dust content of $A_V < 0.15$~mag and a very low metallicity
of $\lesssim 1\%~\Zsun$. Applying a range of magnifications within
the uncertainties would increase or decrease the stellar masses and
SFRs by an additional 15\% or 22\%, respectively. The low metallicity
is consistent with the low dust content and young stellar population
inferred from the SED fitting.

Upcoming JWST Cycle 3 NIRSpec IFU prism observations of the \arc\
will provide a spectroscopic redshift and a definitive test of the
low-metallicity nature of the \arc\ via the \HeIIw\ emission line.
The JWST Cycle 3 observations also include MIRI Medium Resolution
Spectrograph observations that will measure \Ha\ emission, providing
spatially resolved maps of the SFRs and, when combined with the NIRSpec
far-UV spectroscopy, the ionizing photon production efficiency,
$\xi_{\rm ion}$, in a young galaxy 460~Myr after the Big Bang.

\section{Acknowledgments}

Based on observations with the NASA/ESA/CSA JWST obtained from the
Mikulski Archive for Space Telescopes (MAST) at the Space Telescope
Science Institute (STScI), which is operated by the Association of
Universities for Research in Astronomy (AURA), Incorporated, under
NASA contract NAS5-03127. Support for Program number JWST-GO-04212 was
provided through a grant from the STScI under NASA contract NAS5-03127.
The data described here may be obtained from the MAST archive at
\dataset[doi:10.17909/tcje-1780]{https://dx.doi.org/10.17909/tcje-1780}.
Also based on observations made with the NASA/ESA HST, obtained at
STScI, which is operated by AURA under NASA contract NAS5-26555.
The HST observations are associated with programs HST-GO-14096,
HST-GO-15920, HST-GO-9771, HST-GO-12757, and HST-GO-12477. Cloud-based
data processing and file storage for this work is provided by the
AWS Cloud Credits for Research program. The Cosmic Dawn Center is
funded by the Danish National Research Foundation (DNRF) under grant
\#140. A.A. acknowledges support by the Swedish research council
Vetenskapsr{\aa}det (2021-05559). T.H. is supported by the Leading
Initiative for Excellent Young Researchers, MEXT, Japan (HJH02007) and
by JSPS KAKENHI grant No. 22H01258. A.K.I. is supported by JSPS KAKENHI
grant No. 23H00131. M.O. acknowledges the support of JSPS KAKENHI grant
Nos. JP22H01260, JP20H05856, and JP22K21349. Y.T. is supported by JSPS
KAKENHI grant No. 22H04939. R.A.W. acknowledges support from NASA
JWST Interdisciplinary Scientist grants NAG5-12460, NNX14AN10G, and
80NSSC18K0200 from GSFC. A.Z. acknowledges support by grant No. 2020750
from the United States-Israel Binational Science Foundation (BSF) and
grant No. 2109066 from the United States National Science Foundation
(NSF); by the Ministry of Science \& Technology, Israel; and by the
Israel Science Foundation grant No. 864/23. E.V. and M.M. acknowledge
financial support through grants PRIN-MIUR 2020SKSTHZ, the INAF GO Grant
2022 ``The revolution is around the corner: JWST will probe globular
cluster precursors and Population III stellar clusters at cosmic dawn"
and by the European Union--–NextGenerationEU within PRIN 2022 project
No. 20229YBSAN---``Globular clusters in cosmological simulations and in
lensed fields: from their birth to the present epoch."



%

\vspace{5mm}
\facilities{JWST (NIRCam, NIRSpec), HST (ACS, WFC3)}


\software{\astropy\ \citep{astropy2022, astropy2018},
          \photutils\ \citep{photutils111},
          \grizli\ \citep{Grizli},
          \eazypy\ \citep{Brammer2008},
          \bagpipes\ \citep{bagpipes},
          }





\bibliographystyle{aasjournal}



\end{document}